\documentclass[12pt,english]{article}

\usepackage{amssymb}
\usepackage{amsmath}
\usepackage{makecell}
\usepackage{booktabs}
\usepackage{multirow}
\usepackage{enumitem}
\usepackage[linesnumbered,ruled,vlined]{algorithm2e}
\usepackage{natbib}
\usepackage{subcaption}
\usepackage{graphicx}
\usepackage{lmodern}
\usepackage{caption}
\usepackage[table]{xcolor}
\usepackage{array}
\usepackage[english]{babel}
\usepackage[letterpaper,top=2cm,bottom=2cm,left=3cm,right=3cm,marginparwidth=1.75cm]{geometry}
\usepackage[colorlinks=true, allcolors=blue]{hyperref}

\usepackage{authblk}

\title{\fontsize{22}{28}\selectfont A Hierarchical Signal Coordination and Control System Using a Hybrid Model-based and Reinforcement Learning Approach}

\author[a]{Xianyue Peng}
\author[a]{Shenyang Chen}
\author[a,*]{H. Michael Zhang}

\affil[a]{\fontsize{10}{12}\selectfont Department of Civil and Environmental Engineering, University of California, Davis, Davis, CA 95616, United States}

\def\corresemail{hmzhang@ucdavis.edu}

\begin{document}
\maketitle

\footnotemark[1]
\footnotetext[1]{*Corresponding author. Email: \texttt{\corresemail}}

\begin{abstract}
Signal control in urban corridors faces the dual challenge of maintaining arterial traffic progression while adapting to demand variations at local intersections. We propose a hierarchical traffic signal coordination and control scheme that integrates model-based optimization with reinforcement learning. The system consists of: (i) a High-Level Coordinator (HLC) that selects coordination strategies based on observed and predicted demand; (ii) a Corridor Coordinator that derives phase constraints from the selected strategy——either Max-Flow Coordination (MFC) or Green-Wave Coordination (GWC); and (iii) Hybrid Signal Agents (HSAs) that determine signal phases via reinforcement learning with action masking to enforce feasibility. Hierarchical reinforcement learning with Proximal Policy Optimization (PPO) is used to train HSA and HLC policies. At the lower level, three HSA policies——MFC-aware, GWC-aware, and pure agent control (PAC) are trained in conjunction with their respective coordination strategies. At the higher level, the HLC is trained to dynamically switch strategies using a multi-objective reward balancing corridor-level and network-wide performance. The proposed scheme was developed and evaluated on a SUMO–RLlib platform. Case results show that hybrid MFC maximizes throughput under heavy demand; hybrid GWC consistently minimizes arterial stops and maintains progression across diverse traffic conditions but can reduce network-wide efficiency; and PAC improves network-wide travel time in moderate demand but is less effective under heavy demand. The hierarchical design enables adaptive strategy selection, achieving robust performance across all demand levels.
\end{abstract}


\section{Introduction}

Coordinated signal control is widely recognized as an effective strategy to improve corridor traffic performance by facilitating smoother vehicle progression across adjacent intersections. Designing control policies that can adapt to fluctuating traffic demands and diverse system-level objectives remains a significant challenge. Existing dynamic signal control methods can be broadly categorized into three groups: rule-based, model-based, and learning-based. Rule-based methods, including actuated control and backpressure schemes, are typically reactive and decentralized \citep{varaiya2013max, le2015decentralized, ma2020back}. They introduce local flexibility either in phase timing(as seen in actuated control with green extensions) or in phase sequencing (as in backpressure control with dynamic phase selection). These decisions are made based on instantaneous traffic measurements, such as vehicle presence and queue length. Improving
progression between adjacent intersections is achievable under rule-based methods. For example, \cite{cesme2014self} proposed a self-organizing approach that enhances actuated control with secondary extensions and dynamic local coordination. Similarly, Smoothing-MP \citep{xu2024smoothing} extended the backpressure framework by incorporating a smoothing term into the local decision rules, enabling implicit corridor-level coordination without relying on centralized control. However, these methods operate within a reactive decision-making framework that lacks foresight. And their decentralized structure often results in fragile coordination that is susceptible to suboptimal local decisions and poor global performance.

Model-based approaches formulate signal control as an optimization problem guided by mathematical models of traffic flow. A classic example is green-wave coordination, which aligns signal offsets to create progression bands under undersaturated conditions \citep{gartner1991multi, lu2023optimization}. Under heavy demand, queue-constrained models have been proposed to maximize throughput while avoiding downstream spillbacks \citep{michalopoulos1977oversaturated, wu2010identification, hu2013managing, wada2018optimization, wang2022coordinated}. \cite{michalopoulos1977oversaturated} were among the first to analytically address delay minimization under spillback constraints across multiple intersections. Building on this, \cite{wang2022coordinated, peng2023coordinated} developed multi-objective coordinated control models for oversaturated arterials using mixed-integer and quadratic programming techniques. Some model-based methods also incorporate uncertainty, such as the stochastic programming formulation in \cite{tong2015stochastic}, which captures inflow and outflow variability at congested intersections.

While centralized model-based approaches can optimize system-wide performance, they often rely on accurate demand forecasts and static inputs, limiting their robustness in dynamic environments. To enhance adaptability, feedback-based strategies have been developed that utilize real-time traffic measurements to guide signal control. For example, \cite{sun2015quasi} proposed a quasi-dynamic method that adjusts green times based on real-time queue length estimations. Similarly, \cite{ren2016adaptive} introduced an adaptive control scheme that detects potential queue spillbacks using vehicle speed and shockwave analysis, then reallocates green time to prevent intersection blockage. Although these approaches improve local adaptability and reduce reliance on prediction, they are generally limited to one or two intersections, making them difficult to scale up in urban networks.

In contrast, learning-based approaches, especially deep reinforcement learning (RL), provide a data-driven alternative to conventional rule-based and model-based methods. They set up agents to learn optimal policies through trial-and-error interactions without detailed traffic models \citep{wei2021recent, noaeen2022reinforcement}. In traffic signal control applications, RL is typically applied in a decentralized manner, where each intersection acts as an autonomous agent making phase decisions based on local observations to maximize cumulative rewards. Compared to rule-based methods, they can achieve better long-term performance \citep{wei2019colight}. Compared to model-based methods, they allow further flexibility in responding to local traffic conditions.

Despite these benefits, multi-agent coordination remains a core challenge in RL-based signal control \citep{wei2021recent}. In congested scenarios where independent signal agents can hardly capture the gains of coordinated phase patterns, the pure RL-based approaches usually result in suboptimal network-wide performance. To improve coordination, several multi-agent RL (MARL) methods introduce mechanisms such as shared observations or joint policy learning \citep{li2021network, chu2019multi, wang2021adaptive}. For example, \cite{wang2021adaptive} clustered adjacent intersections and employed a shared critic architecture to improve training stability. However, these approaches often increase state space dimensionality and computational complexity, limiting their generalizability and transferability to new environments. Additionally, their performance can be sensitive to hyperparameter tuning, which further complicates real-world deployment \citep{noaeen2022reinforcement}.

Optimizing an urban corridor is a multifaceted task. On one hand, it requires a predictable signal coordination solution to ensure the smooth progression of main street traffic, where model-based approaches excel due to their ability to explicitly optimize offsets and green splits. On the other hand, traffic in other directions often demands greater flexibility, which makes it well-suited for RL-based methods that can adapt to real-time fluctuations and localized conditions. An integrated framework can leverage the strength of both paradigms \citep{sadek2022traffic}. In this regard, we develop a learning-augmented framework that combines model-based coordination policies with multi-agent reinforcement learning techniques to improve the overall performance of the corridor. Specifically, we will cover two coordination policies: 1) a classic Green-Wave Coordination (GWC) policy maximizing the green band for arterial progression; 2) a Max-Flow Coordination (MFC) policy maximizing the arterial throughput. The two policies complement each other under different demand levels: GWC works well under low-to-median demand levels, while MFC works particularly well under high demand levels.

The integrated process is realized by three interdependent components. First, a Corridor Coordinator encodes the policy into a set of phase constraints for each intersection. Then, the Hybrid Signal Agents (HSA), located at each intersection, implement a constrained RL for optimal phase selection among the feasible phase set. The hybrid design enables each agent to adapt to local conditions while maintaining alignment with global control objectives. Finally, a High-Level Coordinator (HLC) is introduced to select the most appropriate schemes based on aggregated congestion and demand levels. The HLC governs the overall control logic by balancing the arterial and network-wide performance. This architecture supports dynamic control logic assignment and enables scalable, efficient signal coordination under diverse traffic conditions.

The main contributions of this paper are as follows:
(1) We propose a hybrid signal control framework that integrates model-based coordination with reinforcement learning, achieving both real-time adaptability and global consistency.
(2) We extend MFC for real-time deployment across multiple intersections, enabling dynamic corridor control.
(3) We introduce a high-level control module that selects coordination strategies based on traffic conditions, improving responsiveness to varying demand levels.

The remainder of this paper is organized as follows. Section~\ref{sec:framework} introduces the overall system architecture. Section~\ref{sec:coordinator} presents the Corridor Coordinator and how it encodes coordination strategies into phase constraints. Section~\ref{sec:hsa} describes the structure and control logic of the Hybrid Signal Agent. Section~\ref{sec:hlc} details the learning mechanism of the High-Level Coordinator. Section~\ref{sec:experiment} reports the simulation results and evaluation. Finally, Section~\ref{sec:conclusion} concludes the paper and outlines future research directions.

\section {Hierarchical Control Architecture}
\label{sec:framework}

\begin{figure}[ht]
\centering
\includegraphics[width=\linewidth]{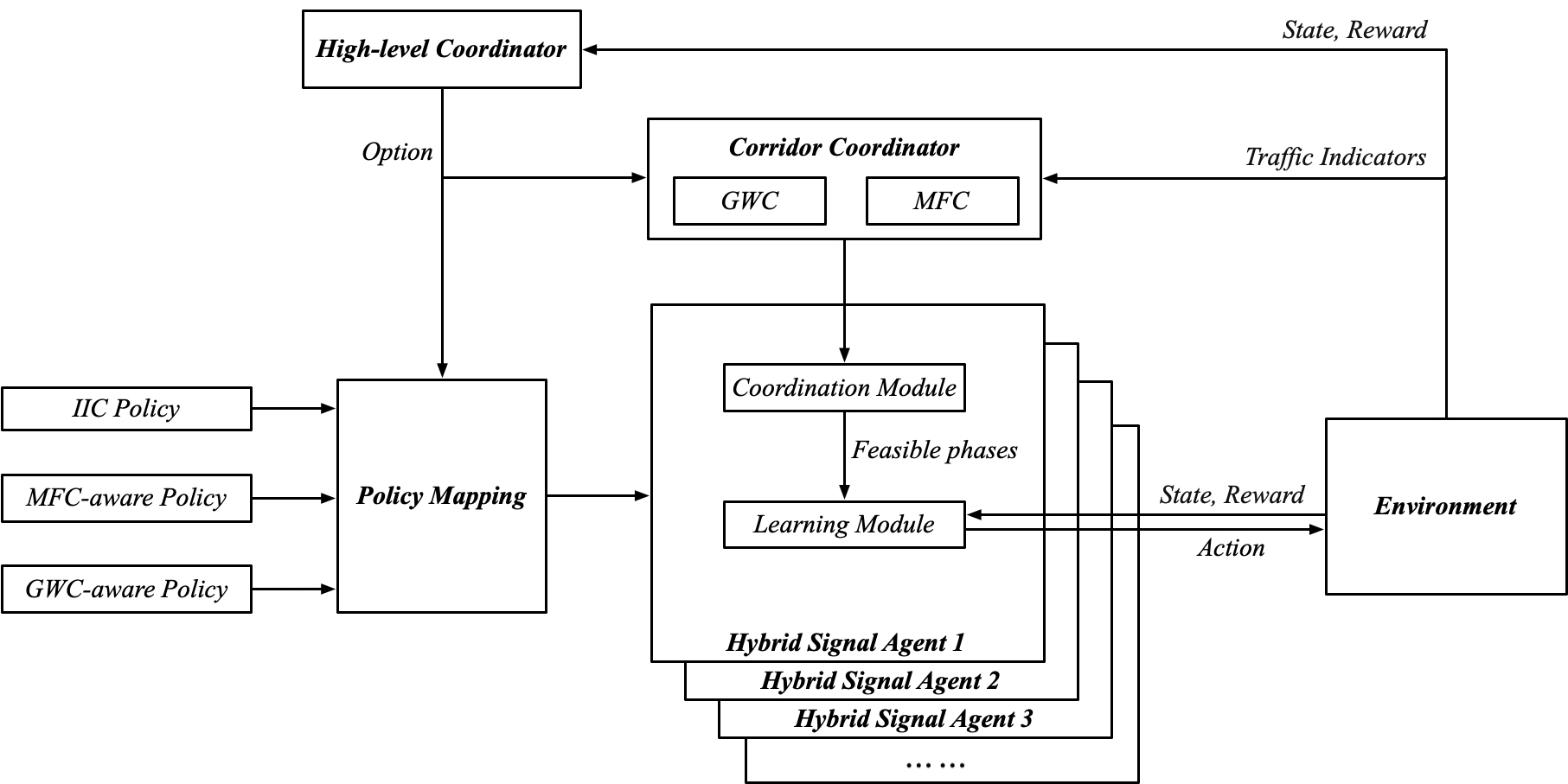} 
\caption{Framework of the Reinforcement Learning Model}
\label{fig:framework}
\end{figure}

To enable adaptive and data-driven coordination of multiple signal control strategies, we design a hierarchical reinforcement learning (HRL) framework. The framework comprises three components: (1) High-Level Coordinator (HLC), (2) Corridor Coordinators, and (3) Hybrid Signal Agents (HSAs) at intersections, as illustrated in Fig.~\ref{fig:framework}.

The High-Level Coordinator selects a coordination strategy \(\omega \in \{\text{PAC}, \text{MFC}, \text{GWC}\}\) based on aggregated traffic states. The selected strategy is then passed to the corresponding Corridor Coordinator, which generates a coordination plan accordingly.

\begin{enumerate}[label=(\arabic*)]
    \item Max-Flow Coordinator: For MFC-managed corridors, inflow, outflow, and vehicle trajectories are modeled, and an MILP is formulated to optimize signal timings during the coordination phase.
    \item Green-Wave Coordinator: For GWC-managed corridors, a green wave model (multi-band model) is applied to compute coordinated signal timings.
\end{enumerate}

Each intersection is controlled by a Hybrid Signal Agent (HSA) capable of making local decisions. Depending on the selected strategy, the HSA operates under one of three policies:
\begin{enumerate}[label=(\arabic*)]
    \item MFC-aware: The agent selects signal phases using a learned policy, with the feasible phase set constrained by the MFC coordination plan via action masking.
    \item GWC-aware: The agent selects phases using a learned policy, constrained by the GWC plan which enforces synchronization of major movements during coordination windows.
    \item Pure Agent Control (PAC): The agent makes decentralized decisions based on locally observed traffic states. To support information sharing while maintaining decentralized control, the state is augmented with arrival flow information from adjacent intersections.
\end{enumerate}

The proposed HRL framework integrates a High-Level Coordinator with multiple low-level HSAs. The high-level option policy assigns a coordination option \(\omega\) to each HSA, which then executes a corresponding intra-option policy. The detailed formulation is presented in Section~\ref{sec:hlc}.

\section{Corridor Coordinator: Model-based coordination}
\label{sec:coordinator}

In this section, we provide the details of the two model-based coordination schemes: Max-Flow Coordination for high demand conditions and Green-Wave Coordination for low to medium demand conditions. First, we introduce the notation used in this section as shown in Table~\ref{tab:model_based_notation}.

\begin{table}[ht]
\centering
\caption{Key Notation Table for the Coordinated Model}
\label{tab:model_based_notation}
\begin{tabular}{ll}
\toprule
\textbf{Symbol} & \textbf{Description} \\
\midrule
\(i\) & intersection index\\
\(k\) & cycle index\\
\(  q^\mathrm{in}_i(k), q^\mathrm{out}_i(k) \)& Inflow and outflow rate at intersection \( i \) during cycle \( k \)\\ 
\( q_{\mathrm{s},i} \)& Saturated flow rate at intersection \( i \)\\ 
\(q_{\mathrm{b},i}\)& Flow rate entering intersection \( i \) from its upstream branch lane.\\ 
\( g_i(k) \)& Green split at intersection \( i \) for the \( k \)-th cycle\\ 
\( \phi_i(k) \)&  Offset for cycle $k$ at intersection \( i \)  \\ 
\(\phi^\mathrm{min}_i(k), \phi^\mathrm{max}_i(k)\) & Minimum and maximum offset for cycle $k$ at intersection \( i \)\\ 
\( C,  z \)& Cycle length and its Inverse value, respectively\\ 
\( L_i \)& Length of edge \( i \)\\ 
\( l_i(k) \)& Queue length at intersection \( i \) during cycle \( k \)\\ 
\(l_i^{\text{max}}(k)\) & Maximum queue length at intersection $i$ within $k$th cycle\\ 
\(t_i\) & Free-flow travel time from upstream intersection \(i-1\) to intersection \(i\) \\
\( t_{\mathrm{c},i}(k) \)& Green split for clearing the residual queue at intersection \(i\) during the \(k\)th cycle \\ 
\( t_{\mathrm{s},i}(k)\)& Green split for saturated flow at intersection \( i \) within the \( k \)th cycle\\ 
\( t_{\mathrm{u},i}(k) \)& Green split for under-saturated flow at intersection \( i \) within the \( k \)th cycle\\ 
\( f_i \)& Proportion of vehicles selecting the optimized direction at intersection \( i \)\\
\( n_i \)& Lane number of coordinated direction at intersection \( i \)\\ 
\( h \)& Stop headway\\ 
\( t_T \) & Total number of cycles in the coordinated control period\\
\bottomrule
\end{tabular}
\end{table}

\subsection{Max-Flow Coordination (MFC)}

To enable signal coordination under high-demand conditions, we propose the MFC method, which aims to maximize corridor throughput and prevent spillback by optimizing flow distribution across intersections.

\textbf{Inflow and Outflow.}  
We define \(q^\mathrm{in}_i(k)\) and \(q^\mathrm{out}_i(k)\) as the through-flow rates entering and exiting intersection \(i\) during cycle \(k\), determined by signal settings and upstream traffic interactions.

The outflow is constrained by either the capacity or total arriving demand:
\begin{equation}
q^\mathrm{out}_i(k) = \min \left(g_i(k)q_{\mathrm{s},i}n_i,\; l_i(k)z + q^\mathrm{in}_i(k) + q_{\mathrm{b},i}(k)\right)
\label{eq:qout}
\end{equation}

The inflow is determined by the upstream outflow and turning ratio:
\begin{equation}
q^\mathrm{in}_i(k) = q^\mathrm{out}_{i-1}(k)f_i
\label{eq:qin}
\end{equation}

\textbf{Queue Dynamics.}  
The queue length is updated based on the incoming demand and the discharge capacity:
\begin{equation}
l_i(k+1) = \max\left(l_i(k) + \frac{\left(q^\mathrm{in}_i(k) + q_{\mathrm{b},i}(k) - g_i(k)q_{\mathrm{s},i}n_i\right)C}{n_i},\; 0\right)
\label{eq:queue}
\end{equation}

The MFC approach formulates mixed-integer linear programs (MILPs) to optimize flow variables across the corridor. Once the optimal inflows and outflows are obtained, signal offsets are calculated to align green phases with vehicle arrivals. The procedure consists of three steps: optimizing the common cycle length \(C\), determining green splits \(g_i(k)\) at each intersection, and computing offsets for phase coordination.

\subsubsection{Cycle Length Optimization}
\label{sec:cycle}
The objective is to maximize the total outflow across all intersections during the first cycle:

\begin{equation}
\max \sum_{i=1}^{n} q^\mathrm{out}_i(1)
\end{equation}
where $q^\mathrm{out}_i(1)$ denotes the outflow rate at intersection $i$ during the first cycle. This step sets $t_T = 1$ and solves the model for the first cycle to determine the optimal cycle length. Given the parameters $L_i$, $l_i(1)$, $f_i$, $q_{\mathrm{s},i}$, $g^\mathrm{max}_i$, $g^\mathrm{min}_i$, $q^\mathrm{max}_{\mathrm{b},i}$, $q^\mathrm{min}_{\mathrm{b},i}$ for all $i \in \{1, 2, \ldots, n\}$, as well as $q^{\mathrm{in}}_1$, $h$, $n$, $C^\mathrm{max}$, and $C^\mathrm{min}$, the objective is to determine the values of $q^\mathrm{out}_i(k)$, $g_i(k)$, $x_i(k)$ for all $i$, $k \in \{1, \ldots, t_T\}$, as well as $q_{\mathrm{b},i}$ and $z$.

The optimization model is subject to the following constraints:
\begin{itemize}
    \item[(C1)] \textit{Outflow rate constraints:} The outflow rate \( q^\mathrm{out}_i(k) \) is bounded by both the green split and the total arriving demand, as shown in Eq.~\ref{eq:qout}. The equivalent linear constraints are:
    \begin{align}
        q^\mathrm{out}_i(k) &\leq g_i(k)q_{\mathrm{s},i}n_i, \\
        q^\mathrm{out}_i(k) &\leq l_i(k)n_iz + q^\mathrm{in}_i(k)+q_{\mathrm{b},i}(k), \\
        q^\mathrm{out}_i(k) &\geq g_i(k)q_{\mathrm{s},i}n_i - M(1-x_i(k)), \\
        q^\mathrm{out}_i(k) &\geq l_i(k)n_iz + q^\mathrm{in}_i(k)+q_{\mathrm{b},i}(k) - Mx_i(k),
    \end{align}
    where \( M \) is a sufficiently large constant and \( x_i(k) \in \{0,1\} \) is a binary variable. When \( x_i(k) = 0 \), the supply (green time) is sufficient to accommodate demand; otherwise, the outflow is limited by the available green time.

    \item[(C2)] \textit{Inflow rate:} The inflow rate \( q^\mathrm{in}_i(k) \) at intersection \( i \) is determined by Eq.~\ref{eq:qin}.

    \item[(C3)] \textit{Spillover avoidance (within current cycle):} The maximum queue length must not exceed the physical storage limit:
    \begin{equation}
        l_i^{\text{max}}(k)h \leq L_i.
    \end{equation}
    According to Fig.~\ref{table:traffic_scenarios}, the maximum queue length is:
    \begin{equation}
        l_i^{\text{max}}(k) = q_{\mathrm{s},i} Ch \max(t_{\mathrm{c},i}(k), t_{\mathrm{s},i}(k)),
    \end{equation}
    where \(t_{\mathrm{c},i}(k)\) and \(t_{\mathrm{s},i}(k)\) respectively denote the green split required to clear the residual queue and the green split for saturated flow at intersection \(i\) in cycle \(k\) (analytical definitions are provided in Eqs.~\ref{eq:tc}–\ref{eq:ts}).

    \item[(C4)] \textit{Green time and branch flow bounds:} The green split must lie within a feasible range:
    \begin{equation}
        g^\mathrm{min}_i \leq g_i(k) \leq g^\mathrm{max}_i,
    \end{equation}
    where \( g^\mathrm{min}_i \) and \( g^\mathrm{max}_i \) are the minimum and maximum allowable green splits.  
    Similarly, the branch flow rate is constrained by:
    \begin{equation}
        q_{\mathrm{b},i}^\mathrm{min} \leq q_{\mathrm{b},i}(k) \leq q_{\mathrm{b},i}^\mathrm{max},
    \end{equation}
    where \( q_{\mathrm{b},i}^\mathrm{min} \) and \( q_{\mathrm{b},i}^\mathrm{max} \) denote the respective bounds.

    \item[(C5)] \textit{Cycle length bounds:} The inverse cycle length \( z = \frac{1}{C} \) must satisfy:
    \begin{equation}
        \frac{1}{C^\mathrm{max}} \leq z \leq \frac{1}{C^\mathrm{min}},
    \end{equation}
    where \( C^\mathrm{min} \) and \( C^\mathrm{max} \) are the minimum and maximum cycle lengths, respectively.
\end{itemize}

\subsubsection{Green Split Optimization}

This step optimizes green splits across multiple cycles to capture both queue evolution and flow variations in the MFC scheme. Unlike GWC, which applies one fixed signal plan over several cycles, MFC must account for residual queues that carry over between cycles; thus, we track \(l_i(k)\) and jointly optimize \(g_i(k)\) for \(k=1,2,\dots,t_T\).
Ideally, both green splits \( g_i(k) \) and cycle length \( C \) (or its inverse \( z \)) would be optimized simultaneously, but doing so over multiple cycles would render the queue dynamic equations, Eq.~\ref{eq:queue}, nonlinear due to the interaction between \( z \) and \( g_i(k) \).
To keep the formulation linear and solvable as an MILP, \( z \) is first optimized in the single-cycle problem described in Section~\ref{sec:cycle} and then fixed here. 
With \( z \) fixed, we introduce state variables \( l_i(k) \) for \( i \in \{1,2,\ldots,n\} \) and \( k \in \{1,2,\ldots,t_T\} \) to model queue dynamics over the horizon.

The goal is to maximize the total outflow across all intersections over the control horizon of \( t_T \) cycles. The objective function is defined as:
\begin{equation}
\max \sum_{k=1}^{t_T} \sum_{i=1}^{n} q^\mathrm{out}_i(k)
\end{equation}

The green split optimization model is subject to Constraints (C1)–(C4) as previously defined, and an additional queue length constraint described below.

\begin{itemize}
    \item[(C6)] \textit{Queue length dynamics:} The queue length at intersection \( i \) evolves according to the balance between inflow and outflow, as defined in Eq.~\ref{eq:queue}. Its equivalent linear form is given by:
    \begin{align}
        l_i(k+1) &\geq l_i(k) + \frac{(q^\mathrm{in}_i(k) + q_{\mathrm{b},i}(k) - g_i(k)q_{\mathrm{s},i}n_i)C}{n_i}, \\
        l_i(k+1) &\geq 0, \\
        l_i(k+1) &\leq l_i(k) + \frac{(q^\mathrm{in}_i(k) + q_{\mathrm{b},i}(k) - g_i(k)q_{\mathrm{s},i}n_i)C}{n_i} + M(1 - x_i(k)), \\
        l_i(k+1) &\leq Mx_i(k),
    \end{align}
    where \( M \) is a sufficiently large constant(Big-M parameter), and \( x_i(k) \in \{0,1\} \) is a binary variable indicating whether demand exceeds supply.
    \item[(C7)] \textit{Spillover avoidance (next cycle):} To prevent queue spillover after the current cycle, the following condition must be satisfied:
    \begin{equation}
        l_i(k+1)h \leq L_i.
    \end{equation}
    By substituting \( z = \frac{1}{C} \), this can be rewritten as:
    \begin{equation}
        l_i(k)z + \frac{q^\mathrm{in}_i(k)+q_{\mathrm{b},i}(k)}{n_i} - g_i(k)q_{\mathrm{s},i} \leq \frac{L_i z}{h}.
    \end{equation}
\end{itemize}

\subsubsection{Offset Calculation}

To facilitate smooth vehicle progression along a coordinated corridor, the offset is adjusted such that inflows arriving from upstream intersections can traverse downstream intersections with minimal delay. The optimal offset depends on the interaction between upstream inflow and downstream outflow dynamics, which may differ significantly under various traffic demand conditions.

\noindent\textbf{Assumption 1.}  
The upstream inflow \(q^\mathrm{in}_i(k)\) is assumed to be homogeneous within each cycle, meaning that vehicles are released at a constant rate during the upstream green interval. This allows the inflow to be represented by a single flow value per cycle.  
In contrast, the downstream outflow \(q^\mathrm{out}_i(k)\) is not assumed to be homogeneous. It typically begins with a high discharge rate during the queue clearance phase and then transitions to a lower, arrival-driven rate. It is further assumed that following vehicles accelerate promptly and close headway gaps once the queue has been discharged.

\noindent\textbf{Assumption 2.}  
Vehicles from branch lanes are assumed to arrive earlier than those from the upstream coordinated approach. This ensures that branch flow is served before the coordinated through-flow arrives and does not interfere with offset computation.

Before introducing the traffic scenarios, it is worth noting that under various traffic demand levels, the discharge flow can exhibit two patterns:  
(1) saturated flow, where vehicles discharge at the maximum flow rate from the start of the green time; and  
(2) under-saturated flow, where vehicles discharge at a lower, arrival-driven rate.  
The corresponding green splits for these patterns, $t_{\mathrm{s},i}(k)$ and $t_{\mathrm{u},i}(k)$, reflect the supply--demand relationship at the intersection.  

We distinguish four traffic scenarios based on two key indicators:  
(1) whether there is residual green time available for the under-saturated flow, i.e., whether $t_{\mathrm{u},i}(k) > 0$; and  
(2) the relationship between the queue clearance time $t_{\mathrm{c},i}(k)$ and the green split for saturated flow $t_{\mathrm{s},i}(k)$, i.e., whether $t_{\mathrm{c},i}(k) \geq t_{\mathrm{s},i}(k)$.  
The analytical definitions of these terms are provided below, and the corresponding scenarios are summarized in Fig.~\ref{table:traffic_scenarios}.

\begin{equation}
    t_{\mathrm{c},i}(k) = \frac{q_{\mathrm{b},i}(k)/n_i + l_i(k)z}{q_{\mathrm{s},i}}
\label{eq:tc}
\end{equation}

\begin{equation}
    t_{\mathrm{s},i}(k) = 
    \begin{cases}
    \frac{q^\mathrm{out}_i(k)}{q_{\mathrm{s},i}n_i} & \text{if } f_i = 1 \\
    \frac{q^\mathrm{out}_i(k)/n_i - g_i(k) f_i q_{\mathrm{s},i}}{q_{\mathrm{s},i}(1 - f_i)} & \text{if } f_i \neq 1
    \end{cases}
\label{eq:ts}
\end{equation}

\begin{equation}
    t_{\mathrm{u},i}(k) = 
    \begin{cases}
    \frac{g_i(k) q_{\mathrm{s},i} - q^\mathrm{out}_i(k)/n_i}{q_{\mathrm{s},i}} & \text{if } f_i = 1 \\
    \frac{g_i(k) q_{\mathrm{s},i} - q^\mathrm{out}_i(k)/n_i}{q_{\mathrm{s},i}(1 - f_i)} & \text{if } f_i \neq 1
    \end{cases}
\end{equation}

\renewcommand{\thetable}{\thefigure} 
\renewcommand{\tablename}{Figure}    
\begin{table}[ht]
\refstepcounter{figure} 
\centering
\setlength{\tabcolsep}{10pt} 
\renewcommand{\arraystretch}{1.5} 
\begin{tabular}{| m{0.45\textwidth} | m{0.45\textwidth} |}
\hline
\multicolumn{2}{|c|}{\cellcolor{gray!25}\textbf{Scenario 1: Partially Saturated Flow at Intersection \( i \) (\( t_{\mathrm{u},i}(k) > 0 \))}} \\ \hline
\centering\includegraphics[width=0.4\textwidth]{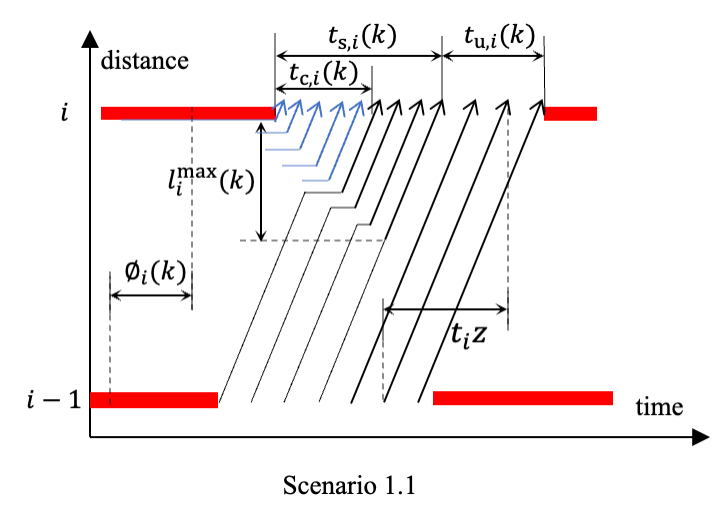} & \centering\includegraphics[width=0.4\textwidth]{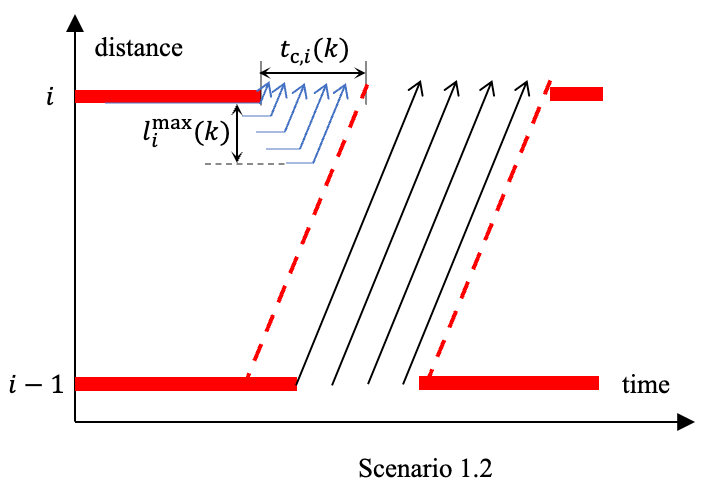} \tabularnewline \hline
\multicolumn{1}{|c|}{\textbf{Scenario 1.1: \( t_{\mathrm{s},i}(k) \geq t_{\mathrm{c},i}(k)\land f_i\neq 1\)}} & \multicolumn{1}{c|}{\textbf{Scenario 1.2: \( t_{\mathrm{s},i}(k) < t_{\mathrm{c},i}(k) \lor f_i =1\)}} \\ \hline
\multicolumn{2}{|c|}{\cellcolor{gray!25}\textbf{Scenario 2: Fully Saturated Flow at Intersection \( i \) (\( t_{\mathrm{u},i}(k) = 0 \))}} \\ \hline
\centering\includegraphics[width=0.4\textwidth]{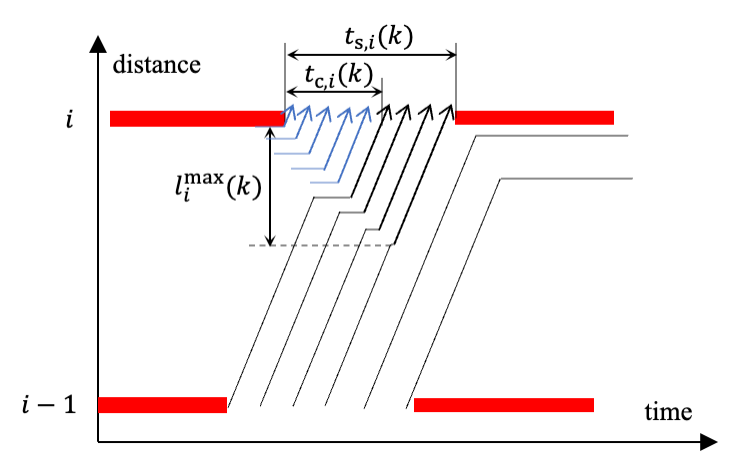} & \centering\includegraphics[width=0.4\textwidth]{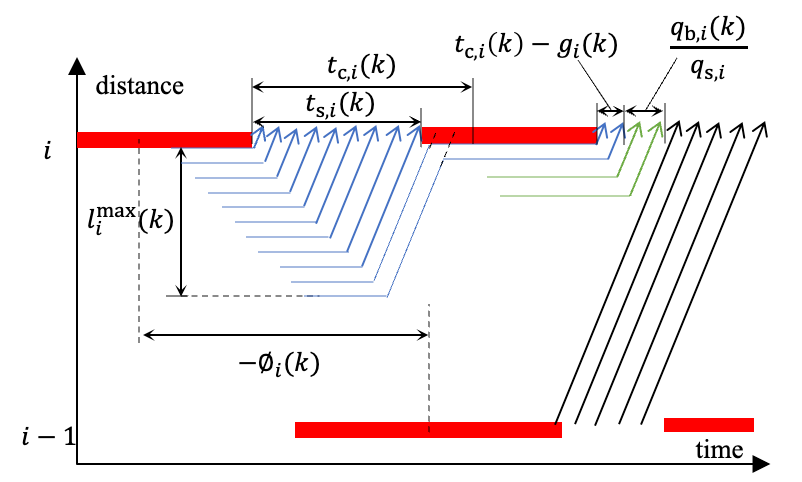} \tabularnewline \hline
\multicolumn{1}{|c|}{\textbf{Scenario 2.1: \( t_{\mathrm{s},i}(k) \geq t_{\mathrm{c},i}(k) \)}} & \multicolumn{1}{c|}{\textbf{Scenario 2.2: \( t_{\mathrm{s},i}(k) < t_{\mathrm{c},i}(k) \)}} \\ \hline
\end{tabular}
\captionsetup{justification=centering}
\caption{Scenarios of Traffic Flow at an Intersection}
\label{table:traffic_scenarios}
\end{table}
\renewcommand{\thetable}{\arabic{table}}
\renewcommand{\tablename}{Table}
\addtocounter{table}{-1} 

\textbf{Scenario 1: Partially Saturated Flow at Intersection \( i \) (\( t_{\mathrm{u},i}(k) > 0 \))}

The outflow through intersection $i$ is not entirely saturated flow; in other words, the flow discharge is at saturation only in part of the green time in each phase. This situation can be further represented as:
\begin{equation}
l_i(k)z + \frac{q_{\mathrm{b},i} + q^\mathrm{in}_i(k)}{n_i} < g_i(k)q_{\mathrm{s},i}
\end{equation}

\textbf{Scenario 1.1: \( t_{\mathrm{s},i}(k) \geq t_{\mathrm{c},i}(k)\land f_i\neq 1\)}

In this scenario, the optimal condition occurs when the last vehicle from the upstream intersection passes through intersection \( i \) as the green time is about to end. 
The optimal offset for achieving this is given by:
\begin{equation}
    \phi_{i}(k) = t_{i} z_i - \frac{g_i(k)}{2}+ \frac{g_{i-1}(k)}{2}  \quad \text{if } i = 2, 3, \ldots, n
\end{equation}
This offset is critical because if it is smaller than the optimal value, vehicles from the upstream intersection may not be able to pass through intersection $i$ in the current cycle. Conversely, if the offset is larger, it results in increased total delay for the upstream flow.

\textbf{Scenario 1.2} $ t_{\mathrm{s},i}(k) < t_{\mathrm{c},i}(k) \lor f_i =1$

The optimal offset for Scenario 1.2 is a range. Importantly, the optimal offset for Scenario 1.1 serves as the lower bound of this range. For Scenario 1.2, the optimal offset can be larger, up to the point where the leading vehicle from the upstream intersection arrives at intersection $i$ precisely when the last vehicle from the queue is about to depart. The range for the optimal offset is defined as:

\begin{equation}
      t_{i} z_i  - \frac{g_i(k)}{2}+ \frac{g_{i-1}(k)}{2} \leq \phi_{i}(k) \leq t_{i} z - t_{\mathrm{c},i}(k) + \frac{g_i(k)}{2} - \frac{g_{i-1}(k)}{2} \quad \text{if } i = 2, 3, \ldots, n
\end{equation}
\begin{equation}
\phi_{i}(k) = t_{i} z_i  - \frac{g_i(k)}{2}+ \frac{g_{i-1}(k)}{2} \quad \text{if } i = 2, 3, \ldots, n
\end{equation}

\textbf{Scenario 2: Fully Saturated Flow at Intersection \( i \) (\( t_{\mathrm{u},i}(k) = 0 \))}

In this scenario, the entire outflow through intersection $i$ is saturated, implying no under-saturated flow. This is mathematically expressed as:
\begin{equation}
l_i(k)z + \frac{q_{\mathrm{b},i} + q^\mathrm{in}_i(k)}{n_i} \geq g_i(k)q_{\mathrm{s},i}
\end{equation}

\textbf{Scenario 2.1} $t_{\mathrm{s},i}(k) \geq t_{\mathrm{c},i}(k)$

In Scenario 1.2, the optimal situation is characterized by two key conditions:
First, the outflow at the intersection should be entirely composed of saturated flow to maximize the overall outflow efficiency.
Second, the upstream flow, which is unable to pass through intersection \( i \) during the current cycle, should only come to a stop after the red signal is activated. This is crucial because if the vehicles stop before the red signal, they will have to stop a second time, leading to increased overall delay and longer queue lengths at the intersection.
The optimal offset is,
\begin{equation}
    \phi_{i}(k) = \left(\frac{1}{f_i} - 1\right) t_{\mathrm{s},i}(k) - \frac{t_{\mathrm{c},i}(k)}{f_i} + \frac{g_i(k)}{2} - \frac{g_{i-1}(k)}{2} + t_{i} z \quad \text{if } i = 2, 3, \ldots, n
\end{equation}

\textbf{Scenario 2.2: \( t_{\mathrm{s},i}(k) < t_{\mathrm{c},i}(k) \)}
In this scenario, the upstream flow is unable to pass through intersection \( i \) within the current cycle. The ideal situation for intersection \( i \) is to set the green split for the upstream intersection \( i-1 \) to 0 within this cycle. This is because allowing the upstream flow to proceed would only lead to a prolonged waiting period without impacting the outflow rate at intersection $i$, regardless of whether there is upstream flow or not. If activating the green time at the upstream intersection \( i-1 \) is necessary, the offset must be carefully calibrated to minimize disruption. 
\begin{equation}
    \phi_{i}(k) = \frac{g_{i}(k)}{2} - \frac{g_{i-1}(k)}{2}+ g_{i}(k+1)+ t_{i} z - t_{\mathrm{c},i}(k)-\frac{q_{\mathrm{b},i}(k) }{q_{\mathrm{s},i}n_i}-1 \quad \text{if } i = 2, 3, \ldots, n
\end{equation}

To summarize, the optimal flow patterns for these scenarios are presented in Fig.~\ref{table:traffic_scenarios}, and the corresponding optimal offset \(\phi^\mathrm{optimal}_i(k)\) for each intersection \(i\) at cycle \(k\) is given as:

\begin{equation}
\phi^\mathrm{optimal}_i(k) =
\begin{cases}
t_{i} z_i - \frac{g_i(k)}{2} + \frac{g_{i-1}(k)}{2} & \text{Scenario 1.1 and 1.2} \\
\left(\frac{1}{f_i} - 1\right)t_{\mathrm{s},i}(k) - \frac{t_{\mathrm{c},i}(k)}{f_i} + \frac{g_i(k)}{2} - \frac{g_{i-1}(k)}{2} + t_{i} z & \text{Scenario 2.1} \\
\frac{g_i(k)}{2} - \frac{g_{i-1}(k)}{2} + g_i(k+1) + t_{i} z - t_{\mathrm{c},i}(k) - \frac{q_{\mathrm{b},i}(k)}{q_{\mathrm{s},i}n_i} - 1 & \text{Scenario 2.2}
\end{cases}
\end{equation}

The final offset is constrained as:
\begin{equation}
\phi_{i}(k) = \max \left( \phi^\mathrm{min}_i(k), \min(\phi^\mathrm{optimal}_i(k), \phi^\mathrm{max}_i(k)) \right)
\end{equation}
where \(\phi^\mathrm{min}_i(k)\) and \(\phi^\mathrm{max}_i(k)\) denote the lower and upper bounds of offset for intersection \(i\) in cycle \(k\), respectively.

\subsection{Green-Wave Coordination (GWC)}  
\label{sec:gwc}

We adopt the classical multi-band model \citep{gartner1991multi} to coordinate signal offsets and green splits for two-way progression. The objective is to maximize the total green bandwidth:
\begin{equation}
\max \sum_{i=2}^{n} (b_i + \overline{b}_i) 
\end{equation}

\begin{figure}[ht]
\centering
\includegraphics[width=0.8\linewidth]{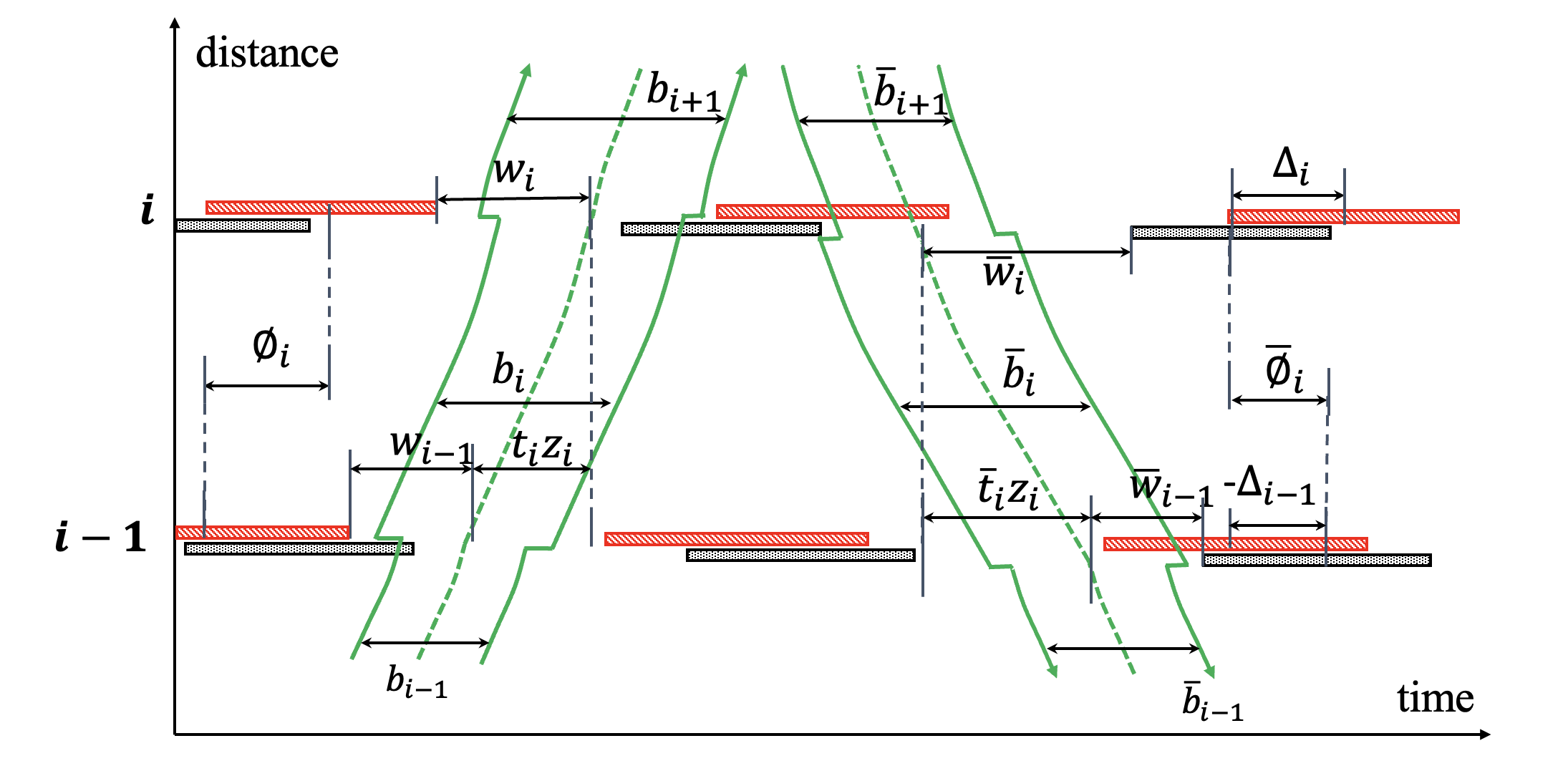} 
\caption{Multi-band Model}
\label{fig: Multi-band model}
\end{figure}

Key variables include green splits \(g_i, \overline{g}_i\), bandwidths \(b_i, \overline{b}_i\), buffers \(w_i, \overline{w}_i\), offsets \(\phi_i, \overline{\phi}_i\), and inverse cycle length \(z = 1/C\), as shown in Fig.~\ref{fig: Multi-band model}. Representative constraints are:
\begin{align}
&w_i + \tfrac{b_i}{2} \leq g_i \\ 
&\overline{w}_i + \tfrac{\overline{b}_i}{2} \leq \overline{g}_i \\
&\phi_i = -\tfrac{g_{i-1}}{2} + \tfrac{g_i}{2} + w_{i-1} - w_i + t_i z \\
&\tfrac{1}{C^\mathrm{max}} \leq z \leq \tfrac{1}{C^\mathrm{min}}
\end{align}
Other model details follow the standard formulation in \citep{gartner1991multi}.

\section{Hybrid Signal Agent: Learning-based Decision Making}
\label{sec:hsa}

Each Hybrid Signal Agent (HSA) controls a signalized intersection and determines the active signal phase at each control step. The agent integrates both model-based coordination and learning-based decision-making to adapt to varying traffic demands and achieve network-level coordination objectives.

\begin{figure}[ht]
\centering
\includegraphics[width=\linewidth]{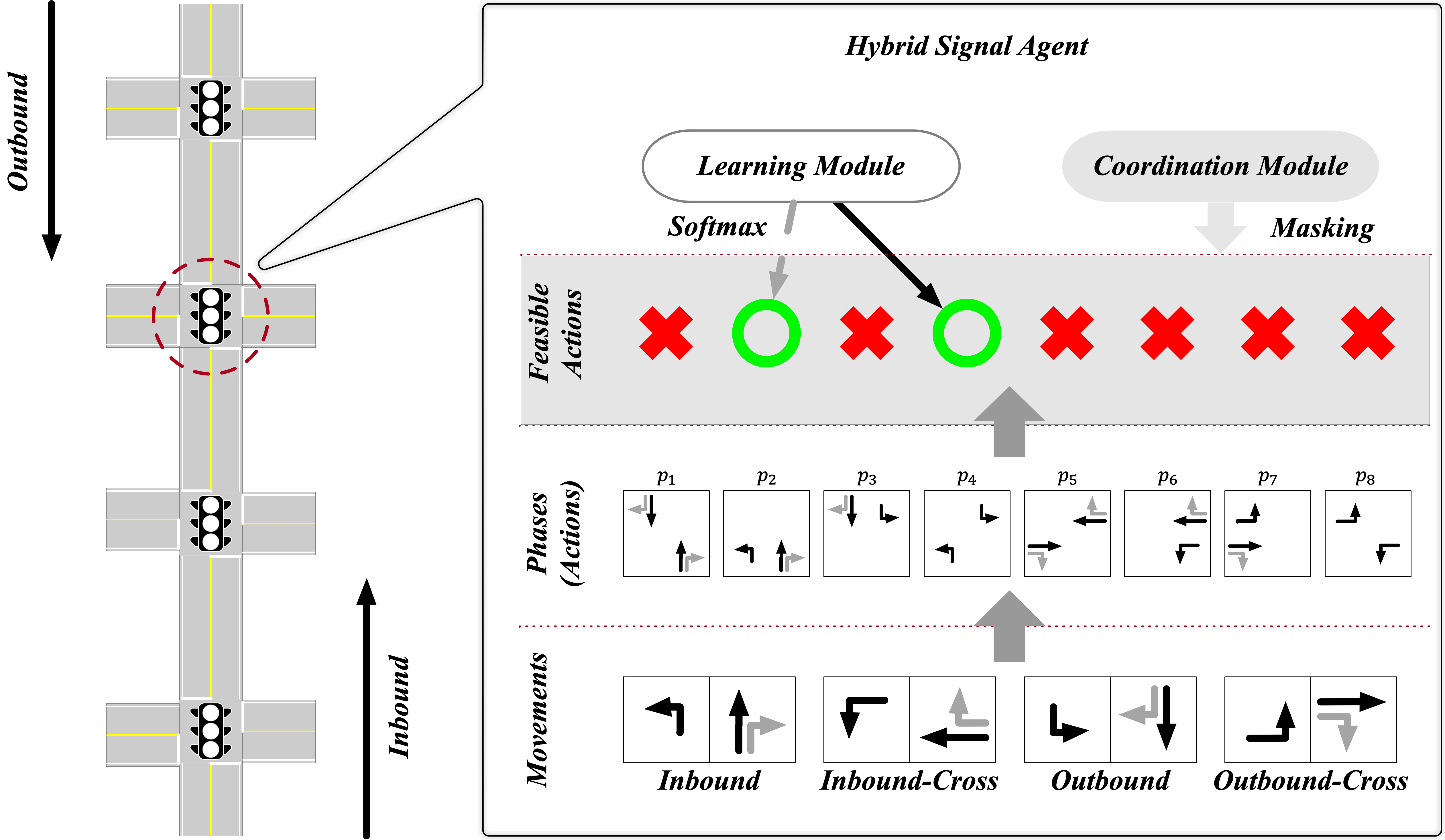} 
\caption{Framework of the Reinforcement Learning Model}
\label{fig:HSA}
\end{figure}

\subsection{Definition in Signalized Intersections}

Taking a typical four-legged intersection as an example, directions are defined relative to the coordinated corridor. Specifically, the four directional groups are inbound, inbound-cross, outbound, and outbound-cross. This movement-based representation is designed to align with coordination strategies, where control decisions are often made along major directional flows. It provides a consistent and structured basis for representing traffic dynamics across intersections.

Based on this directional classification, we define 12 possible movements, among which 8 (including through and left-turn movements) are governed by signal control. Let the set of these controlled movements be denoted as \(M\). Each signal phase enables a subset of these movements while restricting others. Let \(\mathcal{P}_i = \{p_1, p_2, ..., p_8\}\) represent the full set of predefined signal phases, each associated with a specific subset of protected movements.

As illustrated in Fig.~\ref{fig:HSA}, each HSA consists of two main modules:
\begin{itemize}
    \item \textbf{Coordination Module}: Determines a feasible phase set \(P_{i,t} \subseteq \mathcal{P}_i\) based on the current signal coordination strategy (e.g., GWC, MFC, or PAC).
    \item \textbf{Learning Module}: Selects the optimal phase \(a_{i,t} \in P_{i,t}\) at each control step using a constrained reinforcement learning policy.
\end{itemize}

\subsection{Coordination Module}

The coordination module of each HSA restricts the set of feasible signal phases \( P_{i,t} \subseteq \mathcal{P}_i \) based on the global coordination strategy enforced at the intersection. This mechanism allows each agent to align with network-level control objectives while retaining flexibility to make phase decisions within the permitted subset.

Three coordination modes are supported:

\begin{itemize}
    \item \textbf{GWC-aware}: Under the GWC strategy, the coordination module synchronizes major phases (e.g., \(p_1-p_4\)) across corridor intersections to enable bidirectional green bands. During predefined coordination windows, the agent can only choose from these synchronized phases. Outside of these windows, it may select any locally non-conflicting phase.

    \item \textbf{MFC-aware}: Under the MFC strategy, which targets high-demand conditions, coordination focuses on the dominant inbound direction. The coordination module defines a time window and a subset of prioritized phases (e.g., \(\{p_1, p_2\}\)) corresponding to inbound movements. Within this window, the agent selects among the designated phases; outside the window, it may choose from other feasible, non-conflicting phases based on local traffic conditions.

    \item \textbf{PAC}: When no global coordination strategy is enforced, the agent operates in PAC mode. In this case, the coordination module imposes no restrictions, and the agent selects the active phase from the full phase set \( \mathcal{P}_i \), relying solely on local observations.
\end{itemize}

\subsection{Learning Module}
\subsubsection{Policy Design}
Each Hybrid Signal Agent adopts a learning module to select the optimal phase \(a_{i,t} \in P_{i,t}\) from the feasible set determined by the coordination module. To align with the different control strategies applied across the network, we define three reinforcement learning policies, each corresponding to a specific coordination approach: MFC, GWC, and PAC. These policies are denoted as \(\pi_{\theta_{\text{MFC}}}\), \(\pi_{\theta_{\text{GWC}}}\) and \(\pi_{\theta_{\text{PAC}}}\), respectively. All policies are pre-trained and shared across agents to ensure scalability and consistency.

At runtime, each agent selects the appropriate policy through a policy mapping mechanism, which assigns one of the three policies based on the control strategy currently in effect. The selected policy is then used by the learning module to make real-time phase decisions at each control step.

\paragraph{State Observation}

To characterize the traffic condition and dynamics at each intersection, we consider three local features: arrival flow (\( \mu \))—the estimated demand entering the intersection; queue length (\( q \))—the number of stopped or slow-moving vehicles; and average speed (\( \nu \))—the mean velocity of moving vehicles. To enable collaboration among intersections and improve awareness of network conditions, we also include neighbor inflow (\( \xi \))—the arrival flow from adjacent intersections.

For each agent \( i \) at time \( t \), the state vector is defined as:
\begin{equation}
\mathbf{s}_{i,t} = \left( \mathbf{s}^\mu_{i,t},\ \mathbf{s}^q_{i,t},\ \mathbf{s}^\nu_{i,t},\ \mathbf{s}^\xi_{i,t} \right),
\end{equation}
where \( \mathbf{s}^\mu_{i,t},\ \mathbf{s}^q_{i,t},\ \mathbf{s}^\nu_{i,t},\ \mathbf{s}^\xi_{i,t} \in \mathbb{R}^M \) is a movement-based feature vector, and \( M \) denotes the number of controlled movements at intersection \( i \). 

Representing each feature over movements (e.g., inbound through, inbound left-turn,...) enables a unified state structure across heterogeneous intersections, regardless of their geometric layout or demand pattern. Such consistency allows hybrid agents to share model parameters and facilitates efficient multi-agent learning and coordination.

\begin{table}[ht]
\centering
\caption{Notation Table for Agent-based Control}
\begin{tabular}{ll}
\toprule
\textbf{Symbol} & \textbf{Description} \\
\midrule
$i$ & Index of intersection (also agent index) \\
$t$ & Index of low-level control step for HSA\\
$\tau$ & Index of high-level decision step for HLC\\
$U_i$ & Set of approach lanes at intersection $i$ \\
$N$ & Set of intersections along the corridor \\
$M$ & Set of movements at each intersection \\
$\mathbf{s}_\tau$ & State observed by the High-Level Coordinator at step $\tau$ \\
$\mathbf{s}^d_\tau$ & Predicted demand level for the upcoming control period \\
$\mathbf{s}^q_\tau$ & Vector of average queue lengths on inbound through lanes at step $\tau$\\
$\mathbf{s}^{\bar{q}}_\tau$ & Vector of average queue lengths on outbound through lanes at step $\tau$\\
$\mathbf{s}_{i,t}$ & State vector observed by agent $i$ at time $t$ \\
$\mathbf{s}^\mu_{i,t}$ & Movement-based arrival flow vector at intersection $i$ and time $t$ \\
$\mathbf{s}^q_{i,t}$ & Movement-based vector of queue lengths at intersection $i$ and time $t$ \\
$\mathbf{s}^\nu_{i,t}$ & Movement-based average speed vector at intersection $i$ and time $t$ \\
$\mathbf{s}^\xi_{i,t}$ & Movement-based inflow from adjacent intersections at intersection $i$ and time $t$ \\
$r_\tau$ & High-level reward at step $\tau$ \\
$r^s_\tau$ & Total vehicle stops along the corridor during step $\tau$ \\
$r^v_\tau$ & Average vehicle speed along the corridor during step $\tau$ \\
$r_{i,t}$ & Reward of agent $i$ at time $t$ \\
$r^q_{i,t,u}$ & Queue lengths in lane $u$ at intersection $i$ and time $t$ \\
$r^{w}_{i,t,u}$ & Waiting time of the first vehicle in lane $u$ at intersection $i$ and time $t$\\
$\Omega$ & Set of predefined coordination strategies: \{PAC, MFC, GWC\} \\
$\omega_\tau$ & Coordination strategy selected at step $\tau$ \\
$a_{i,t}$ & Action taken by agent $i$ at time $t$ \\
$P_{i,t}$ & Feasible phase set of intersection $i$ at control step $t$ \\
$\pi_{\theta}(\omega_\tau \mid \mathbf{s}_\tau)$ & High-level policy parameterized by $\theta$ \\
$\pi_{\theta_\omega}(a_i \mid \mathbf{s}_i)$ & Low-level policy for policy $\omega$, parameterized by $\theta_\omega$ \\
\bottomrule
\end{tabular}
\label{tab:agent_control_notation}
\end{table}

\paragraph{Action Selection with Action Masking}

The action for each agent $i$ at time $t$ is selected from a constrained action set $P_{i,t} \subseteq \mathcal{P}_i$. We apply an action mask \citep{huang2020closer} to enforce feasibility:

\begin{equation}
a_{i,t} \in P_{i,t},\quad \text{via masked } \pi_{\theta_\omega}(a_{i,t} \mid \mathbf{s}_{i,t}).
\end{equation}

To implement this, we define a binary mask vector 
\begin{equation}
\mathbf{m}_{i,t} = (m_{i,t,1}, m_{i,t,2}, \dots, m_{i,t,|\mathcal{P}_i|}) \in \{0,1\}^{|\mathcal{P}_i|}
\end{equation}
as the indicator vector of $P_{i,t}$, where each entry corresponds to an action $a \in \mathcal{P}_i$:
\begin{equation}
m_{i,t,a} = \mathbf{1}[a \in P_{i,t}].
\end{equation}

Let $\mathbf{z}_{i,t}$ be the unmasked logits output by the policy network for agent $i$, from which the original (unmasked) policy is defined as:
\begin{equation}
\pi^{\text{unmasked}}_{\theta_{\omega}}(a_{i,t} \mid \mathbf{s}_{i,t}) = \text{Softmax}(\mathbf{z}_{i,t})_{a_{i,t}} = \frac{e^{z_{i,t,a_{i,t}}}}{\sum_{a' \in \mathcal{P}_i} e^{z_{i,t,a'}}}.
\end{equation}

We apply a masked transformation to the logits:
\begin{equation}
z'_{i,t,a} = \begin{cases}
z_{i,t,a} & \text{if } m_{i,t,a} = 1 \\
-\infty & \text{if } m_{i,t,a} = 0
\end{cases}
\quad \text{or equivalently} \quad \mathbf{z}'_{i,t} = \mathbf{z}_{i,t} + \log \mathbf{m}_{i,t}.
\end{equation}

The resulting masked policy becomes:
\begin{equation}
\pi_{\theta_\omega}(a_{i,t} \mid \mathbf{s}_{i,t}) = \text{Softmax}(\mathbf{z}_{i,t} + \log \mathbf{m}_{i,t})_{a_{i,t}},
\end{equation}

ensuring that $\pi_{\theta_\omega}(a_{i,t} \mid \mathbf{s}_{i,t}) = 0$ for all infeasible actions $a_{i,t} \notin P_{i,t}$.

\paragraph{Reward Design}
We design the reward by considering both queue length and head vehicle delay, following \citep{chu2019multi}. Specifically, the reward for agent $i$ at time $t$ is defined as:
\begin{equation}
\displaystyle r_{i,t} = -\sum_{u \in U_i} \left( r^{q}_{i,t,u} + \kappa r^{w}_{i,t,u} \right)
\end{equation}
where the coefficient $\kappa$ balances the two objectives.

\subsubsection{PPO-Based Policy Optimization}
\label{sec:ppo}
For each coordination mode \(\omega \in \{\text{PAC}, \text{MFC}, \text{GWC}\}\), a single policy \(\pi_{\theta_\omega}\) and value function \(V_{\sigma_\omega}\) is shared across all agents operating under that mode. These shared policies are trained using the Proximal Policy Optimization (PPO) algorithm based on data collected from all relevant agents, enabling consistent and scalable learning across the network.

At each training iteration, the advantage function is estimated as:
\begin{equation}
\hat{A}_{i,t}^{(\omega)} = r_{i,t} + \gamma \cdot V_{\sigma_\omega}(\mathbf{s}_{i,t+1}) - V_{\sigma_\omega}(\mathbf{s}_{i,t}),
\end{equation}
where \(r_{i,t}\) is the immediate reward computed from the environment, \(\gamma\) is the discount factor, and \(V_{\sigma_\omega}(\cdot)\) is the value function associated with policy \(\pi_{\theta_\omega}\), parameterized by $\sigma_\omega$.

The policy update is performed by maximizing the clipped surrogate objective:
\begin{equation}
J^{(\omega)}(\theta_\omega) = \mathbb{E} \left[ \min\left( \rho^{(\omega)}_{i,t} \hat{A}^{(\omega)}_{i,t},\ \text{clip}(\rho^{(\omega)}_{i,t}, 1 - \epsilon, 1 + \epsilon) \hat{A}^{(\omega)}_{i,t} \right) \right],
\end{equation}
where $\epsilon$ is a clipping parameter that constrains the policy update to stay within a trust region, and $\rho^{(\omega)}_{i,t}$ is the probability ratio defined as
\begin{equation}
\rho^{(\omega)}_{i,t} = \frac{\pi_{\theta_\omega}(a_{i,t}|\mathbf{s}_{i,t})}{\pi_{\theta'_\omega}(a_{i,t}|\mathbf{s}_{i,t})},
\end{equation}
with $\pi_{\theta_\omega}$ and $\pi_{\theta'_\omega}$ denoting the current and previous policy $\omega$, respectively.

The value function is optimized by minimizing the temporal-difference (TD) error:
\begin{equation}
L^{(\omega)}(\sigma_\omega) = \mathbb{E} \left[ \left( r_{i,t} + \gamma V_{\sigma_\omega}(\mathbf{s}_{i,t+1}) - V_{\sigma_\omega}(\mathbf{s}_{i,t}) \right)^2 \right].
\end{equation}

These policy and value function updates are applied independently for each coordination strategy \(\omega\), based on samples collected from agents operating under that strategy. This training paradigm allows each policy to specialize in a distinct traffic control pattern, while maintaining scalability through parameter sharing within each mode.

\section{High-Level Coordinator}
\label{sec:hlc}

The proposed HRL framework follows a two-level agent design~\citep{pateria2021hierarchical}, which integrates high-level decision-making over coordination strategies (referred to as options) with low-level control over signal phases (known as intra-option policies)~\citep{bacon2017option}. Each option lasts for a predefined duration \(\tau\), which spans a sequence of lower-level control steps \(t\).

At the high level, the High-Level Coordinator (HLC) adopts a policy \(\pi_\theta(\omega_\tau \mid \mathbf{s}_\tau)\) that selects a joint option \(\omega_\tau \in \Omega = \{\text{PAC}, \text{MFC}, \text{GWC}\}\) based on the aggregated corridor-level state \(\mathbf{s}_\tau\). Each option specifies the coordination strategy applied to all intersections in the corridor and remains fixed for the duration \(\tau\) before a new decision is made.

At the low level, each intersection \(i\) is controlled by a Hybrid Signal Agent (HSA), which executes an intra-option policy \(\pi_{\theta_{\omega}}(a_i \mid \mathbf{s}_i)\). Here, \(\mathbf{s}_i\) denotes the local traffic state at intersection \(i\), and \(\theta_{\omega}\) represents the policy parameters associated with the selected high-level strategy \(\omega\). Each HSA consists of a coordination module that constrains the feasible phase set according to the assigned strategy \(\omega\), and a learning module that selects a phase within that constrained set.

\subsection{Policy Design}

\paragraph{State Observation}

To characterize the traffic condition along the coordinated corridor, we construct the high-level observation vector to incorporate both temporal demand indicators and spatial features of the coordinated movements.

At each high-level decision step \( \tau \), the observed state is defined as:
\begin{equation}
\mathbf{s}_\tau = \left( \mathbf{s}^d_{t}, \mathbf{s}^q_{\tau}, \mathbf{s}^{\bar{q}}_{\tau} \right),
\end{equation}
where the state incorporates the predicted demand level for the upcoming coordination period, as well as the average queue lengths of inbound through and outbound through movements.

\paragraph{Option Selection}

The HLC selects a coordination strategy \(\omega_\tau \in \Omega \). Once selected, the same strategy is synchronously applied to all intersections and remains active for the duration of step \( \tau \):
\begin{equation}
 \omega_\tau \in \Omega ,\quad \text{via } \pi_\theta(\omega_\tau \mid \mathbf{s}_\tau).
\end{equation}

\paragraph{Reward Design}

The high-level reward is designed to assess the effectiveness of the selected coordination strategy by considering both network-level congestion and corridor-level mobility. It includes three components: (i) the total queue length across the entire network during step $\tau$, reflecting overall traffic accumulation; (ii) the total number of vehicle stops along the coordinated path, indicating the smoothness of flow in prioritized directions; and (iii) the average vehicle speed along the coordinated path, representing the progression efficiency. The reward is computed as:

\begin{equation}
r_\tau = -\alpha_1 \sum_{t \in \tau} \sum_{i \in N} \sum_{u \in U_i} r^{q}_{i,t,u} - \alpha_2 r^s_\tau + \alpha_3 r^v_\tau,
\end{equation}
where \( \alpha_1, \alpha_2, \alpha_3\) are coefficients for weighting reward components.

\subsection{Hierarchical Reinforcement Learning Algorithm}

The HLC's policy is updated using the Proximal Policy Optimization (PPO) algorithm, as described in Section~\ref{sec:ppo}. The high-level policy \(\pi_\theta(\omega_\tau \mid \mathbf{s}_\tau)\), parameterized by \(\theta\), is optimized via the clipped surrogate objective:
\begin{equation}
\label{eq:hlc_policy_obj}
J(\theta) = \mathbb{E} \left[ \min\left( \rho_\tau \hat{A}_\tau,\ \text{clip}(\rho_\tau, 1 - \epsilon, 1 + \epsilon) \hat{A}_\tau \right) \right],
\end{equation}
where the advantage estimate \(\hat{A}_\tau\) for high-level step \(\tau\) is computed as
\(\hat{A}_\tau = r_\tau + \gamma V_{\sigma}(\mathbf{s}_{\tau+1}) - V_{\sigma}(\mathbf{s}_\tau)\), and the probability ratio is defined as
\(\rho_\tau = \frac{\pi_{\theta}(\omega_\tau \mid \mathbf{s}_\tau)}{\pi_{\theta'}(\omega_\tau \mid \mathbf{s}_\tau)}\), with \(\theta'\) denoting the parameters of the previous policy.

The value function \(V_{\sigma}(\cdot)\), parameterized by \(\sigma\), is updated by minimizing the temporal-difference error:
\begin{equation}
\label{eq:hlc_value_loss}
L(\sigma) = \mathbb{E} \left[ \left( r_\tau + \gamma V_{\sigma}(\mathbf{s}_{\tau+1}) - V_{\sigma}(\mathbf{s}_\tau) \right)^2 \right].
\end{equation}

The training procedure for the High-Level Coordinator is summarized in Algorithm~\ref{alg:hlc_ppo}. Each intra-option policy \(\pi_{\theta_{\omega}}\) is pre-trained and kept fixed throughout the high-level training, allowing the coordinator to learn effective corridor-level strategies without interfering with low-level control adaptation.

\begin{algorithm}[ht]
\caption{High-Level Coordinator Training with PPO}
\label{alg:hlc_ppo}
\SetAlgoLined
\KwIn{Trained intra-option policies $\pi_{\theta_\omega}$ for $\omega \in \{\text{PAC}, \text{MFC}, \text{GWC}\}$}
\KwOut{Updated high-level policy $\pi_{\theta}$ and value function $V_{\sigma}$}
\BlankLine
Initialize high-level policy $\pi_{\theta}$ and value function $V_{\sigma}$\;
\For{each PPO training iteration}{
    Initialize empty buffer $\mathcal{D} \leftarrow \emptyset$\;
    \For{each rollout episode}{
        Reset environment\;
        \For{each high-level step $\tau$}{
    Observe state $\mathbf{s}_{\tau}$\;
    Sample option $\omega_{\tau} \sim \pi_{\theta}(\cdot \mid \mathbf{s}_{\tau})$\;
    \For{$t \in \tau$}{
        Run low-level agents using frozen $\pi_{\theta_\omega}$\;
    }
    Compute reward $r_{\tau}$, observe $\mathbf{s}_{\tau+1}$\;
    Store $(\mathbf{s}_{\tau}, \omega_{\tau}, r_{\tau}, \mathbf{s}_{\tau+1})$ into $\mathcal{D}$\;
}
    }
    \For{each transition in $\mathcal{D}$}{
        Estimate advantage: $\hat{A}_\tau = r_\tau + \gamma V_{\sigma}(\mathbf{s}_{\tau+1}) - V_{\sigma}(\mathbf{s}_\tau)$\;
    }
    \For{each PPO epoch}{
        \ForEach{minibatch $(\mathbf{s}_{\tau}, \omega_{\tau}, \hat{A}_{\tau}) \in \mathcal{D}$}{
            Compute probability ratio: $\rho_\tau = \frac{\pi_{\theta}(\omega_\tau \mid \mathbf{s}_\tau)}{\pi_{\theta'}(\omega_\tau \mid \mathbf{s}_\tau)}$\;
            Update $\pi_{\theta}$ by maximizing PPO objective (Eq.~\ref{eq:hlc_policy_obj})\;
            Update $V_{\sigma}$ by minimizing TD loss (Eq.~\ref{eq:hlc_value_loss})\;
        }
    }
    Update old policy: $\theta' \leftarrow \theta$\;
}
\end{algorithm}

\section{Case Study}
\label{sec:experiment}
\subsection{Experimental Setup}

The simulation is conducted in SUMO \citep{SUMO2018}, interfaced with TraCI in synchronous mode. A custom multi-agent environment, built on the RLlib framework \citep{liang2018rllib}, enables hierarchical policy updates and coordination between high-level and low-level agents.

We consider an arterial corridor consisting of six signalized intersections. Three traffic demand scenarios (low, medium, and high) are used to evaluate the adaptability and robustness of the proposed methods. Detailed simulation settings and training code are available at: \url{https://github.com/xypeng12/hierarchical-signal-rl-mpc}.

\subsubsection{Training Setup and Hyperparameters}

The lower-level Hybrid Signal Agents (HSAs) are trained separately under three demand levels (low, medium, high). Each training episode lasts 3600 seconds, including a 600-second warm-up and a 600-second evaluation window for collecting metrics. The demand level is randomly selected at the beginning of each episode.

The High-Level Coordinator (HLC) is trained with longer episodes, each simulating 16 hours of traffic flow from 5:40 to 22:00, during which the traffic demand varies dynamically to reflect realistic temporal patterns. The high-level control step is 3600 seconds, while the low-level step duration is 3 seconds. At the start of each high-level step, a 600-second measurement phase is included to collect stable traffic parameters for coordination. During this period, the PAC control strategy is enforced at the lower level.

All policies are trained using the PPO algorithm with entropy regularization and action masking. Table~\ref{tab:hyperparameters} summarizes the key hyperparameters and network architecture settings for both levels.

\begin{table}[htbp]
\centering
\caption{Training Hyperparameters and Network Architectures}
\label{tab:hyperparameters}
\begin{tabular}{lll}
\hline
\textbf{Parameter} & \textbf{Low-Level HSA} & \textbf{High-Level Coordinator} \\
\hline
Episode duration & 3600 s & 58800 s (16 hr 20 min) \\
Warm-up period & 600 s & 1200 s \\
Control step & 3 s & 3600 s \\
Train batch size per learner & 20000 & 20000 \\
Minibatch size & 1024 & 1024 \\
Number of PPO epochs & 20 & 20 \\
Clipping parameter (\(\epsilon\)) & 0.3 & 0.3 \\
KL divergence coefficient & 0.2 & 0.2 \\
Value function clip & 1000.0 & 1000.0 \\
Entropy coefficient & 0.005 & 0.005 \\
Learning rate schedule & \((0, 5\text{e}{-4})\),\newline \((2\text{e}{5}, 1\text{e}{-4})\),\newline \((5\text{e}{5}, 1\text{e}{-5})\) & same as left \\
Network hidden layers & [256, 128] & [256, 128] \\
Activation function & ReLU & ReLU \\
Training iterations & 300 & 30 \\
\hline
\end{tabular}
\end{table}

\subsubsection{Metrics}

To evaluate both the overall performance of each control strategy and the learning progress during training, we track a set of episode-level metrics. These metrics are designed to capture travel efficiency (e.g., travel time and throughput), coordination effects (e.g., throughput, stop frequency and average speed along the corridor), and agent-level decision quality (e.g., total reward).

\begin{itemize}
\item \textbf{Total travel time (TT)}:
The total travel time is reported separately for inbound through (InTT), outbound through (OutTT), and other uncoordinated movements (OthTT).

\item \textbf{Average travel time (AvgT)}:  
The average time each vehicle spends in the network, calculated as the total travel time divided by the number of vehicles.

\item \textbf{Throughput (Thru)}:  
The total number of vehicles, all vehicles within the network and along the corridor that reach their destinations during an episode.

\item \textbf{Average stops along the corridor (Stop)}:  
The average number of stops experienced by vehicles traveling along the corridor (both inbound and outbound direction).

\item \textbf{Average speed along the corridor (Speed)}:  
The average speed of vehicles traveling along the corridor (both inbound and outbound direction).

\item \textbf{Total reward}:  
The sum of rewards accumulated by all agents during the episode, which reflects the control performance in terms of queue lengths and vehicle waiting times.

\end{itemize}

\subsection{Hybrid model-based and learning-based control}

\subsubsection{Training Process}

Fig.~\ref{fig:learning_curves} shows that all three strategies generally achieve good convergence, with most performance metrics stabilizing as training continues. Under high demand conditions, the system is more prone to congestion, making control stability especially important. Compared to others, MFC exhibits smoother and more stable performance during training, indicating stronger robustness. In contrast, although PAC has strong potential, its performance fluctuates noticeably at times, suggesting that poor decisions under high demand may trigger rapid congestion and a drop in performance.

As the demand level decreases, the advantages of PAC become more apparent. Its flexible control logic, which does not rely on fixed coordination windows, allows it to respond dynamically to vehicle arrivals. This adaptability leads to significantly improved performance across multiple metrics—particularly in terms of average travel time and efficiency for non-coordinated movements. MFC continues to demonstrate balanced and stable performance, while GWC retains its advantage in supporting smooth flow along the arterial, especially in minimizing stops.

In the early training stages, both MFC and GWC are effective in improving arterial flow, suggesting that coordination-based strategies can begin to take effect immediately. Even before full optimization, they can improve flow along the corridor. Non-coordinated strategies start with weaker performance on the arterial but gradually enhance flow across other movements as training progresses, eventually showing stronger network-wide adaptability.

In summary, under high-demand scenarios, coordination-based strategies—particularly MFC—offer more consistently effective control. In contrast, under medium and low demand conditions, PAC demonstrates greater flexibility and the ability to optimize overall network performance.

\begin{figure}[htbp]
    \centering

    \begin{subfigure}{\textwidth}
        \includegraphics[width=\linewidth]{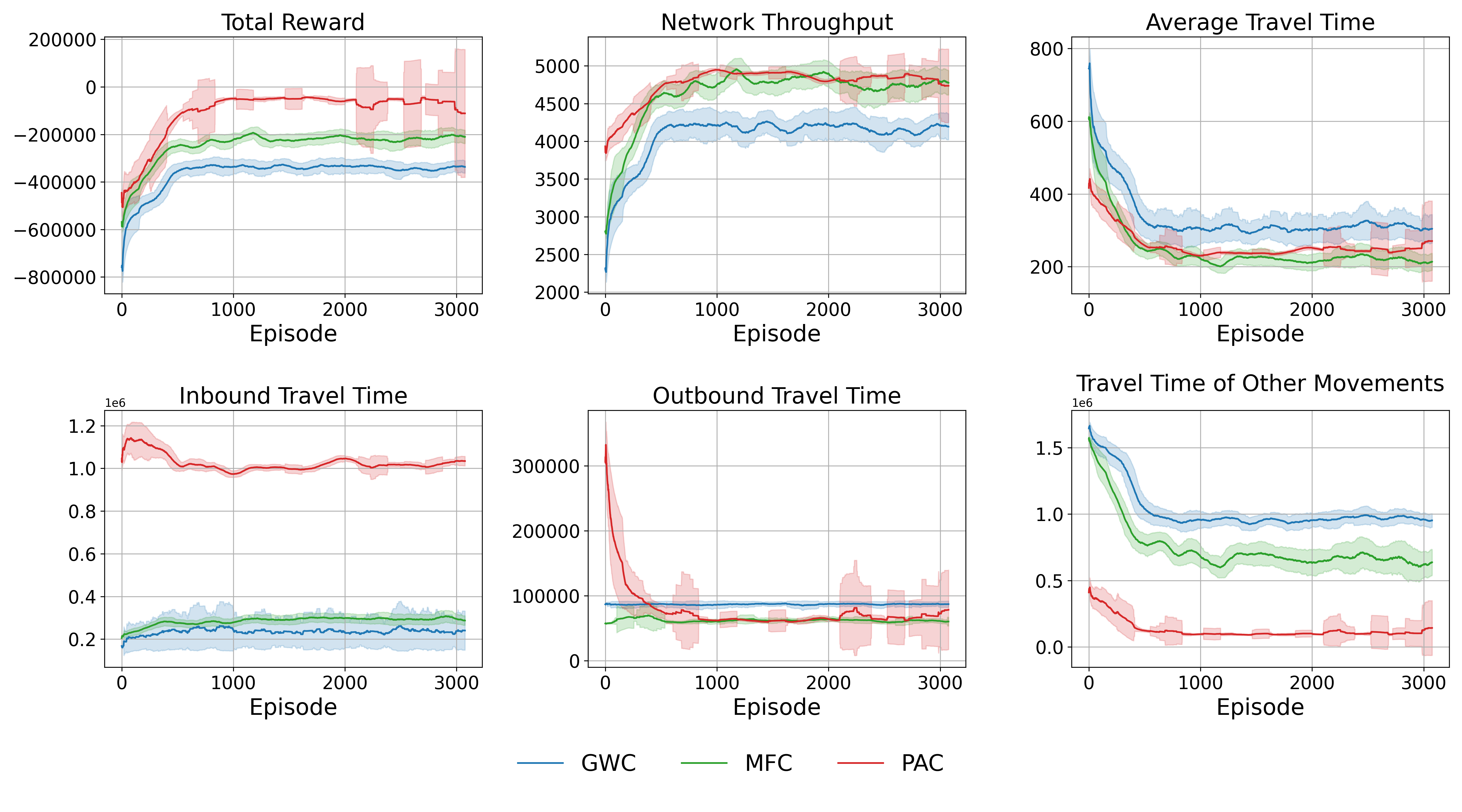}
        \caption{High demand}
        \label{fig:learning_high}
    \end{subfigure}
    
    \vspace{0.4em}

    \begin{subfigure}{\textwidth}
        \includegraphics[width=\linewidth]{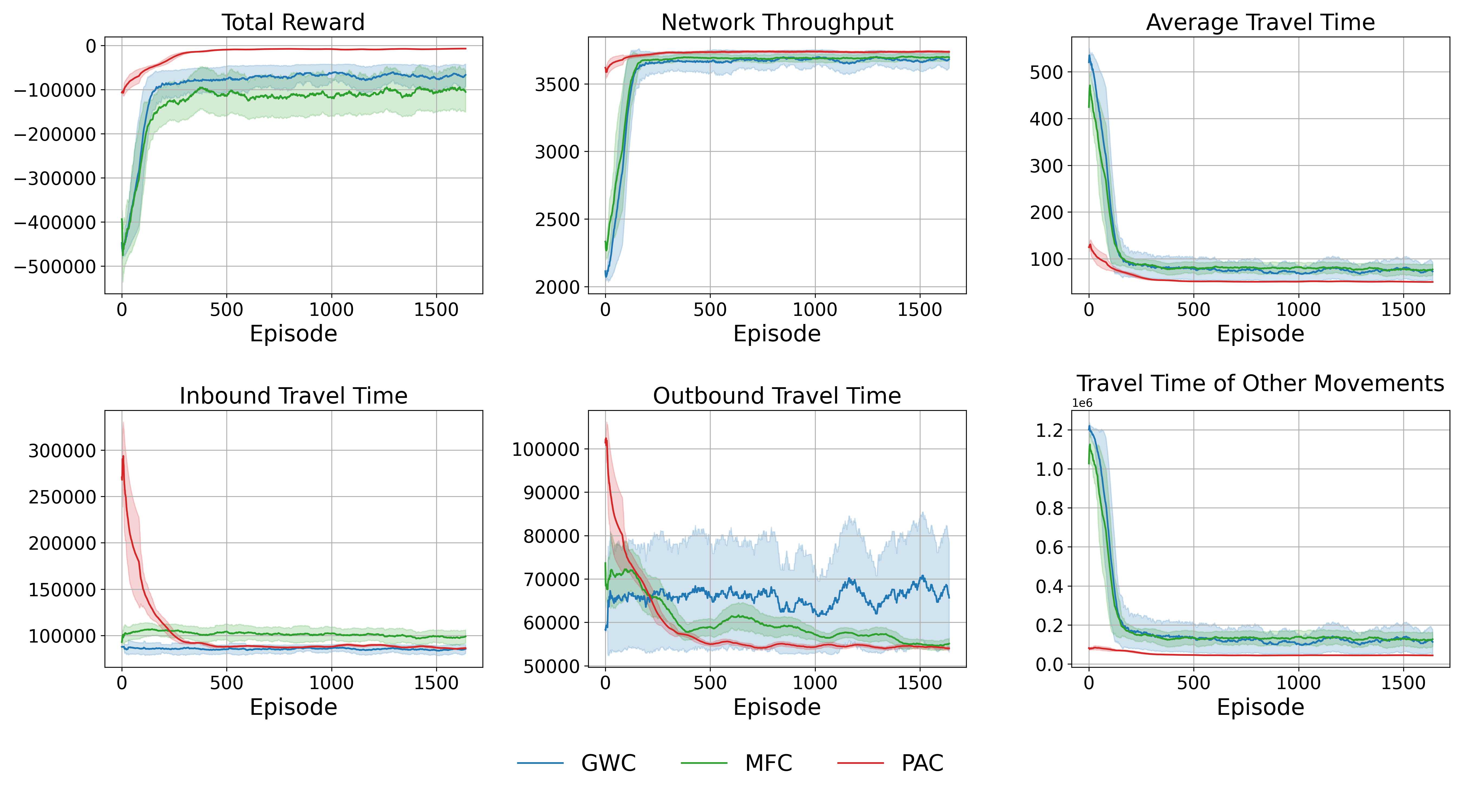}
        \caption{Medium demand}
        \label{fig:learning_medium}
    \end{subfigure}
    
    \caption{Learning curves under different demand levels. Each subplot summarizes training progress across six performance metrics.}
    \label{fig:learning_curves}
\end{figure}

\begin{figure}[htbp]
    \ContinuedFloat
    \centering

    \begin{subfigure}{\textwidth}
        \includegraphics[width=\linewidth]{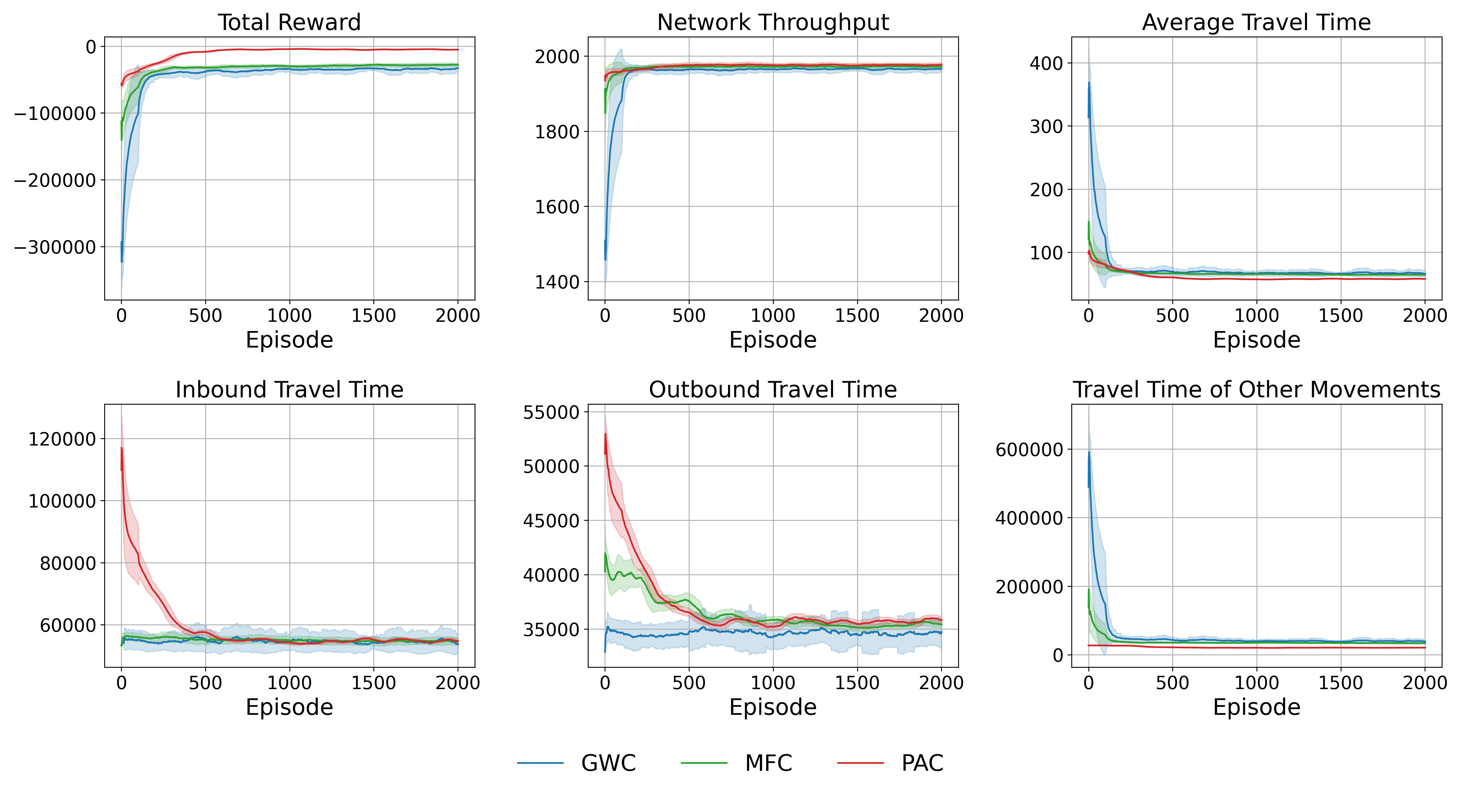}
        \caption{Low demand}
        \label{fig:learning_low}
    \end{subfigure}
    
    \caption{(Continued) Learning curves under different demand levels.}
\end{figure}
\subsubsection{Policy Evaluation}

We implemented the three trained policies in the simulation, and compared them against the Backpressure (BP) baseline \citep{varaiya2013max}. The evaluation considers eight episode-level metrics introduced in Section~5.1.3, and the results are summarized in Table~\ref{tab:metrics-summary}. 

MFC demonstrates superior performance in improving arterial throughput under high demand. It achieves the highest corridor throughput (2216 vehicles), significantly outperforming GWC (2116 vehicles), PAC (2027 vehicles), and BP (1839 vehicles), while maintaining an average corridor speed of 10.09~m/s. 

Despite favoring the corridor flow, MFC does not increase overall travel time; instead, its average travel time (183.61~s) is substantially lower than that of GWC (236.30~s). These results suggest that by explicitly modeling inflow and outflow rates, MFC is able to allocate green times more efficiently, fully utilizing available capacity without sacrificing non-coordinated movements. In contrast, GWC focuses on maximizing green-band progression and often allocates the longest possible green time to the corridor direction without explicitly accounting for actual flow rates, which can negatively impact the performance of non-coordinated movements.

GWC excels at maintaining uninterrupted flow along the corridor by minimizing stop frequency and improving speed. Across all demand levels, GWC consistently yields the lowest average number of stops (no more than 0.76) and the highest average speed (no less than 10.31~m/s). This strategy is especially effective when demand is moderate and the corridor has a clear directional priority. 

PAC performs well in optimizing network-wide travel time under low and medium demand. For example, in the medium-demand scenario, PAC achieves an average travel time of 50.88~s, compared to 62.54~s for GWC, 87.98~s for MFC and 63.41~s for BP. This is likely attributed to PAC's distributed, local optimization mechanism, which balances movements based on pressure differences without prioritizing any specific corridor. 

While PAC performs well under low and medium demand, its lack of coordination becomes a major drawback under high demand, resulting in more stops and longer travel times (Table~\ref{tab:metrics-summary}). To better understand this, we examine space-time trajectories along the inbound corridor Fig.~\ref{tab:trajectory-grid-hybrid}. Under PAC, vehicles exhibit dense stop-and-go patterns, with frequent deceleration across intersections and upstream-propagating congestion waves. In contrast, coordinated strategies—especially GWC—enable most vehicles to pass with at most one stop, confirming the importance of coordination in sustaining smooth flow.

\begin{table}
\centering
\caption{Performance metrics under different strategies and demand levels (learning)}
\label{tab:metrics-summary}
\begin{tabular}{@{} l r r r  r r r r r r}
\toprule
& & \multicolumn{3}{c}{\textbf{Corridor metrics}} & \multicolumn{5}{c}{\textbf{Network-wide metrics}} \\
\cmidrule(lr){3-5} \cmidrule(lr){6-10}
\textbf{Dem.} & \textbf{Str.} & \textbf{Thru} & \textbf{Stop} & \textbf{Speed} & \textbf{Thru} & \textbf{AvgT} & \textbf{InTT} & \textbf{OutTT} & \textbf{OthTT} \\
\midrule
\multirow{4}{*}{High} 
& MFC & 2216 & 1.09 & 10.09 & 5087 & 183.61 & 309{,}786 & 368{,}463 & 521{,}838 \\
& GWC & 2116 & 0.76 & 10.31 & 4619 & 236.30 & 229{,}206 & 320{,}547 & 746{,}811 \\
& PAC & 2072 & 1.50 & 4.36  & 4874 & 241.25 & 1{,}024{,}746 & 1{,}084{,}098 & 86{,}934 \\
& BP  & 1839 & 5.64 & 2.88  & 4536 & 328.28 & 1{,}109{,}637 & 107{,}955 & 319{,}749 \\
\midrule
\multirow{4}{*}{Medium} 
& MFC & 1127 & 0.61 & 12.54 & 3659 & 87.98 & 100{,}353 & 158{,}175 & 160{,}839 \\
& GWC & 1134 & 0.45 & 13.16 & 3716 & 62.54 & 87{,}183 & 146{,}223 & 80{,}550 \\
& PAC & 1131 & 0.54 & 13.26 & 3729 & 50.88 & 87{,}066 & 140{,}964 & 43{,}527 \\
& BP  & 1148 & 2.02 & 10.57 & 3729 & 63.41  & 111{,}135 & 66{,}189  & 52{,}587 \\
\midrule
\multirow{4}{*}{Low} 
& MFC & 747 & 0.55 & 13.75 & 1981 & 63.58 & 54{,}717 & 90{,}147 & 32{,}373 \\
& GWC & 733 & 0.19 & 14.58 & 1967 & 64.89 & 54{,}849 & 88{,}269 & 36{,}774 \\
& PAC & 744 & 0.52 & 13.68 & 1988 & 57.38 & 55{,}467 & 91{,}026 & 19{,}554 \\
& BP  & 742 & 1.99 & 11.57 & 1965 & 70.19  & 64{,}821  & 42{,}867  & 27{,}576 \\
\bottomrule
\end{tabular}
\end{table}

\begin{table}[htb]
\centering
\renewcommand{\thetable}{\thefigure} 
\renewcommand{\tablename}{Figure}    
\refstepcounter{figure} 
\setcounter{table}{3}  

\renewcommand{\arraystretch}{1.2}
\begin{tabular}{
  >{\centering\arraybackslash}m{1.2cm}  
  >{\centering\arraybackslash}m{0.28\textwidth}
  >{\centering\arraybackslash}m{0.28\textwidth}
  >{\centering\arraybackslash}m{0.28\textwidth}
}

& \textbf{High} & \textbf{Medium} & \textbf{Low} \\

\textbf{MFC} &
\includegraphics[width=\linewidth]{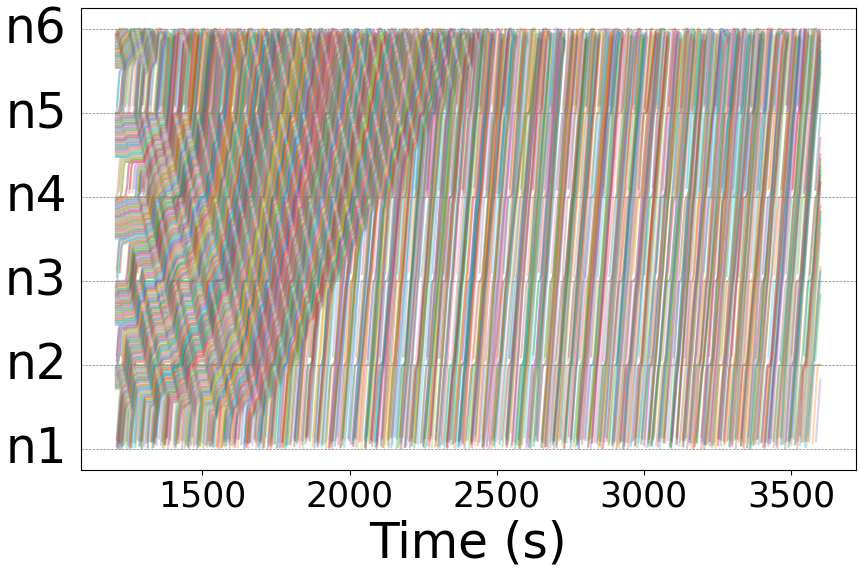} &
\includegraphics[width=\linewidth]{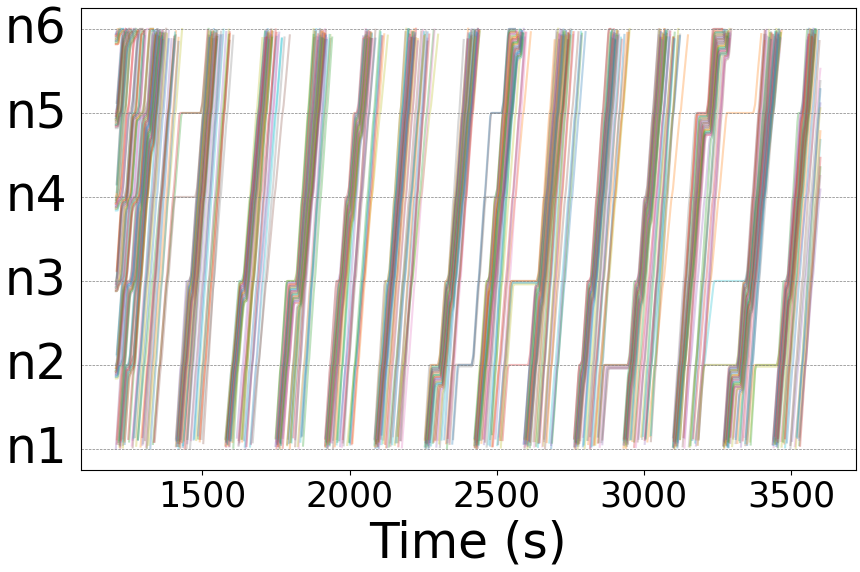} &
\includegraphics[width=\linewidth]{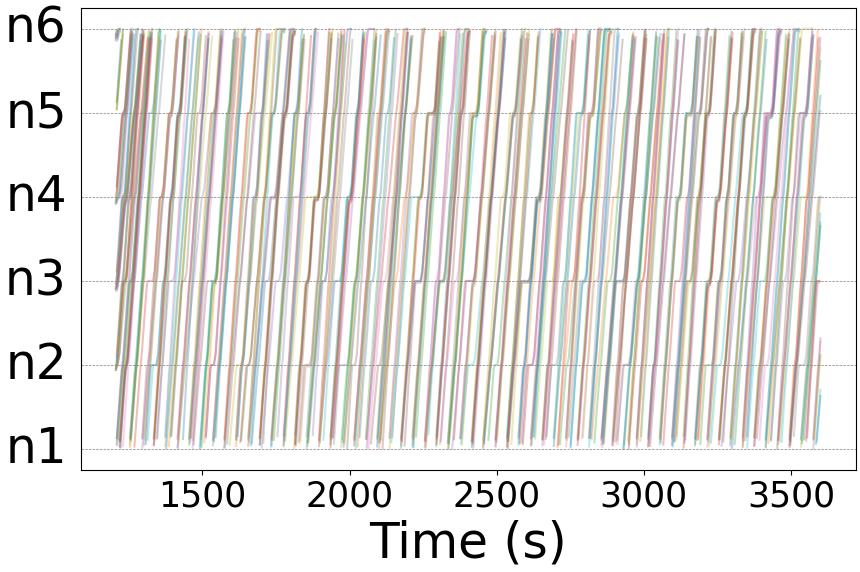}
\\

\textbf{GWC} &
\includegraphics[width=\linewidth]{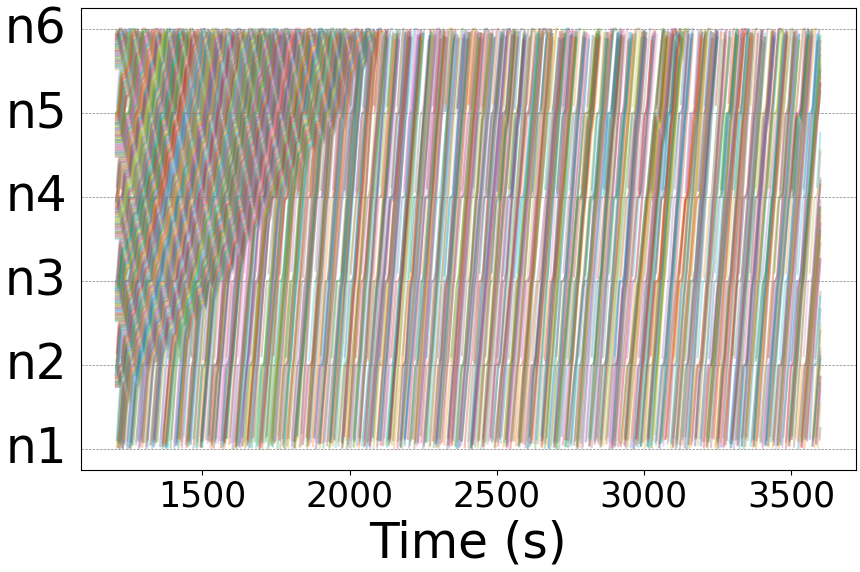} &
\includegraphics[width=\linewidth]{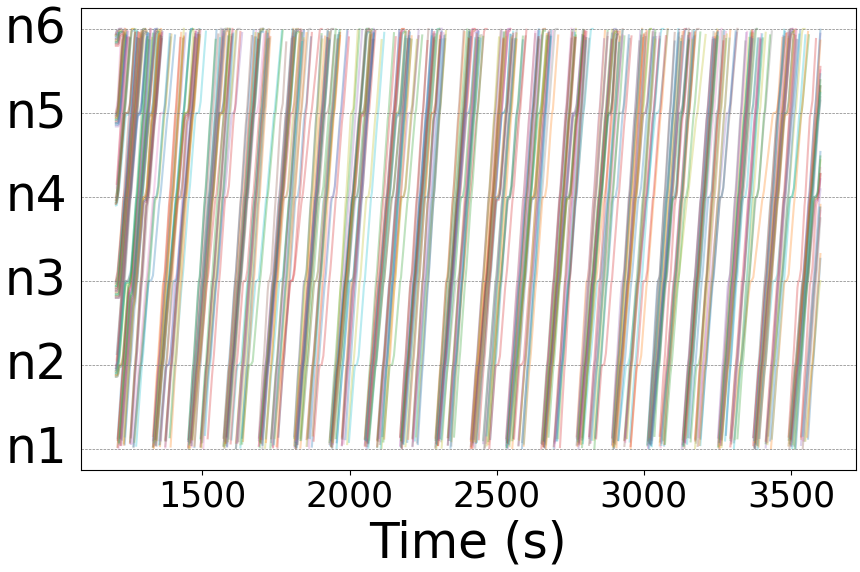} &
\includegraphics[width=\linewidth]{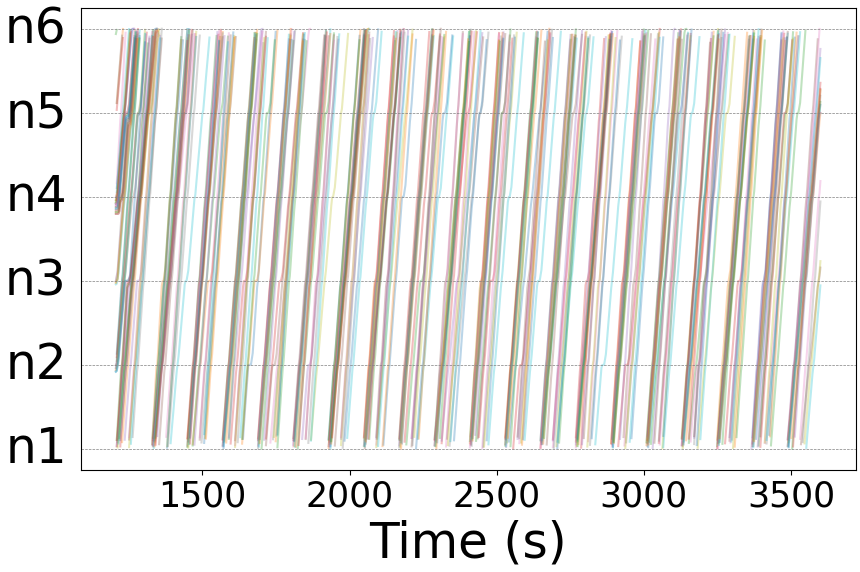}
\\

\textbf{PAC} &
\includegraphics[width=\linewidth]{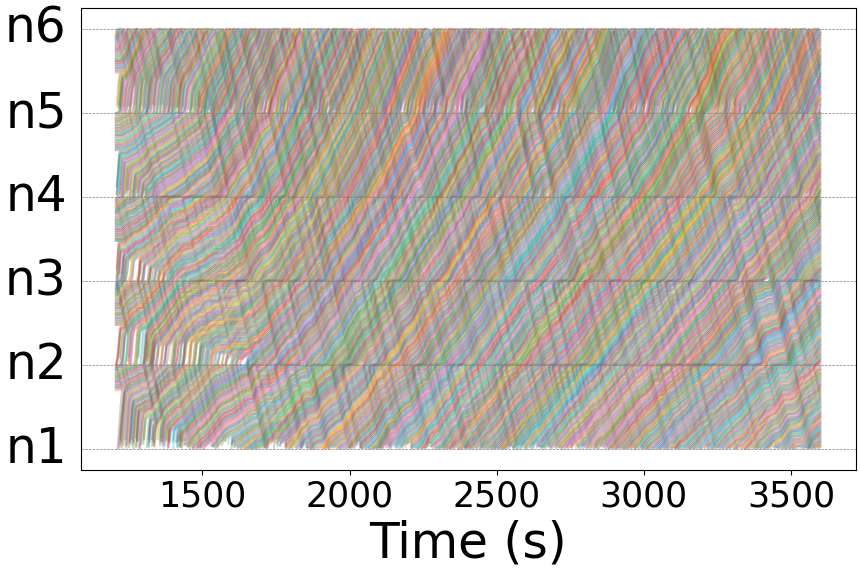} &
\includegraphics[width=\linewidth]{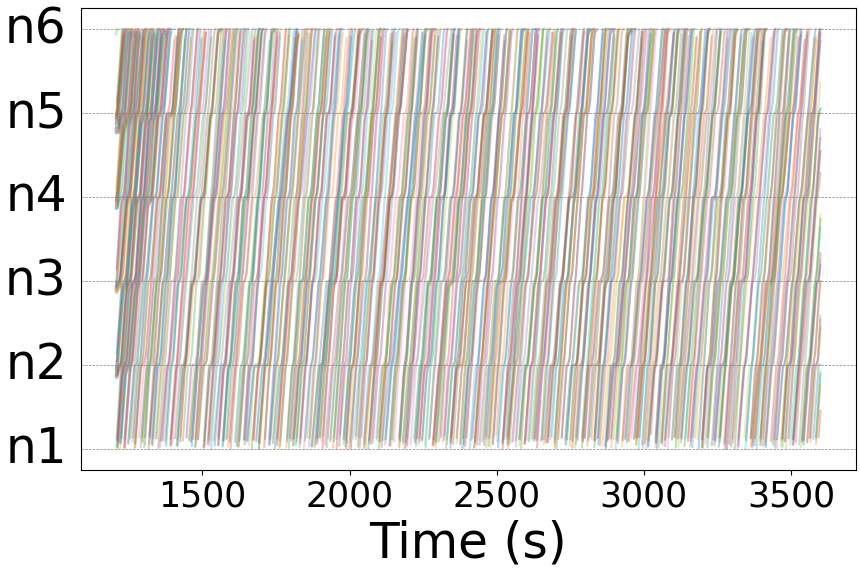} &
\includegraphics[width=\linewidth]{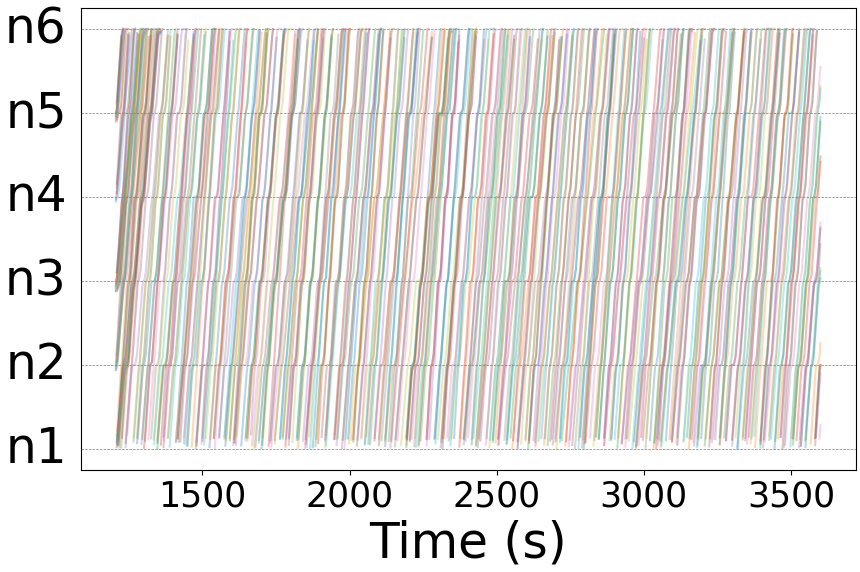}
\\

\textbf{BP} &
\includegraphics[width=\linewidth]{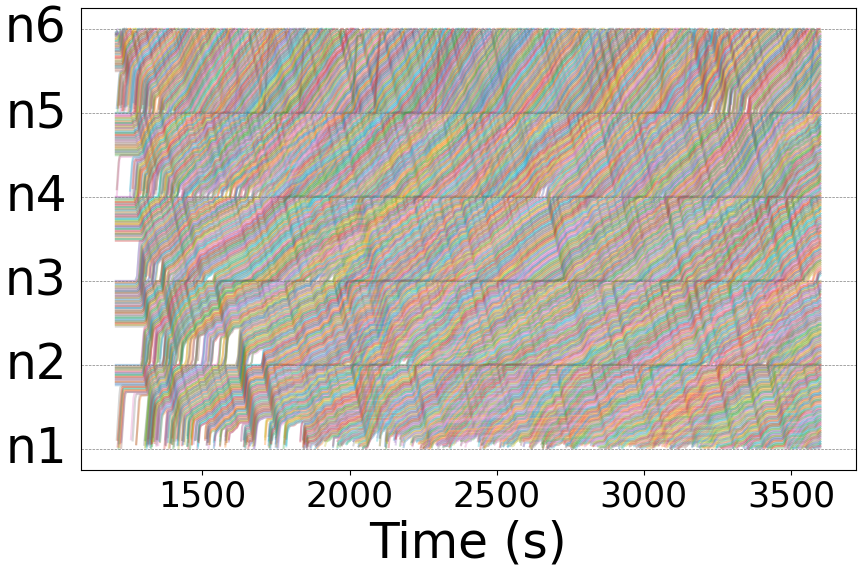} &
\includegraphics[width=\linewidth]{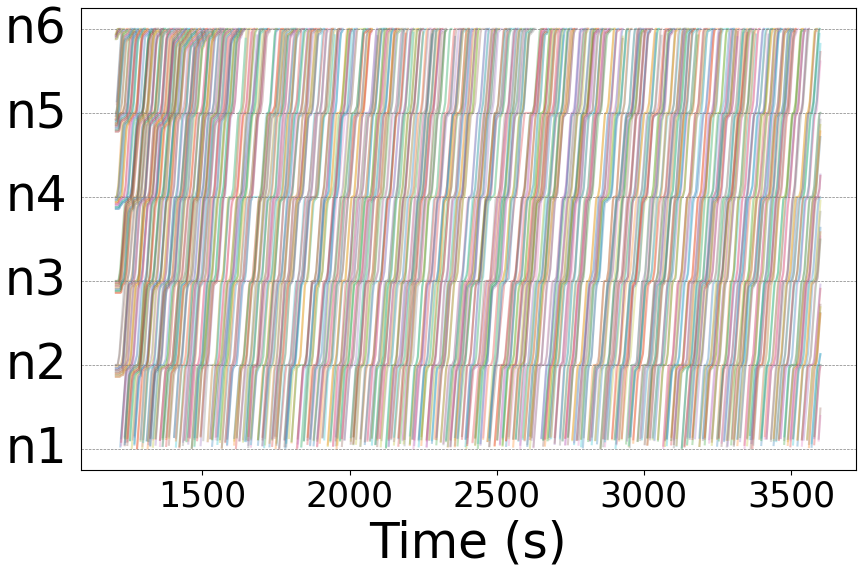} &
\includegraphics[width=\linewidth]{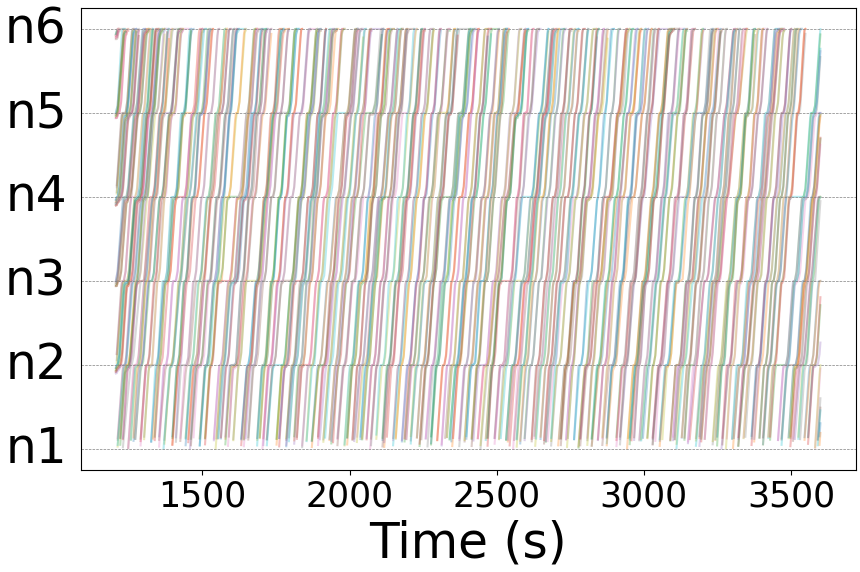}
\\

\end{tabular}
\caption{Trajectory visualizations under different strategies (rows) and demand levels (columns). }
\label{tab:trajectory-grid-hybrid}
\end{table}

It is also worth noting that under high-demand scenarios, coordinated plans often require a brief warm-up period before benefits are fully realized. Although coordination is activated at a specific time (e.g., 1200s), the downstream arrival patterns shortly afterward are still shaped by upstream signal plans applied before coordination began. It takes some time for the updated upstream signal timings to influence the vehicle flow reaching downstream intersections. Once this propagation takes effect and arrival patterns become aligned with the coordinated scheme, queues are gradually cleared and throughput improves. Nevertheless, even during this early transition phase, coordinated strategies still outperform PAC and BP by mitigating shockwaves and promoting smoother flow.

Overall, both MFC and GWC outperform PAC and BP in terms of corridor-level performance, while PAC demonstrates clear advantages in network-wide metrics under low and medium demand conditions. MFC is particularly effective under high demand, benefiting from its flow-aware green time allocation, whereas GWC is better suited for moderate demand scenarios where minimizing stops is a primary objective. PAC consistently outperforms BP across all demand levels, owing to its responsiveness to neighboring intersection states. These findings highlight the complementary strengths of MFC, GWC, and PAC under varying demand conditions and control objectives.

\subsection{Hierarchical Multi-policy}

The reward structure in the high-level controller (HLC) defines the intended coordination objective. By adjusting the weights associated with corridor performance and network-wide stability, we shape the HLC’s preference toward different traffic management goals. To investigate this influence, we train the HLC under three configurations:

\begin{itemize}
    \item Group 1: \((\alpha_1, \alpha_2, \alpha_3) = (-1, -0.1, 50)\)
    \item Group 2: \((\alpha_1, \alpha_2, \alpha_3) = (-1, -0.01, 10)\)
    \item Group 3: \((\alpha_1, \alpha_2, \alpha_3) = (-1, -0.005, 1)\)
\end{itemize}
Here, \(\alpha_1\), \(\alpha_2\), and \(\alpha_3\) penalize network-wide queue length, encourage average speed, and penalize arterial stops along the coordination path, respectively. These groups reflect increasing emphasis on global stability (Group 3) versus corridor throughput (Group 1).

\begin{figure}[ht]
\centering
\includegraphics[width=\linewidth]{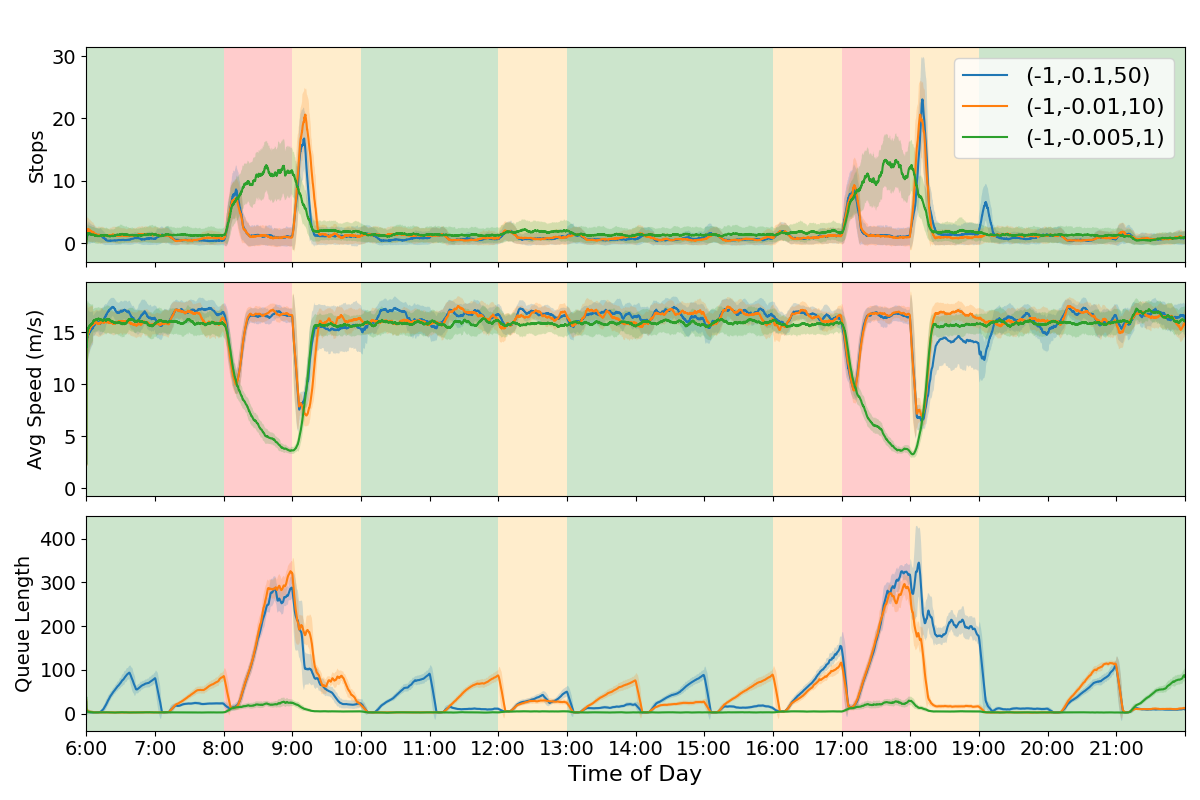}
\caption{Performance under different high-level reward settings. }
\label{fig:hl_param}
\end{figure}

The trained policies are evaluated under a shared demand profile to examine how HLC adapts to different traffic states. As shown in Fig.~\ref{fig:hl_param}, we compare stops, average speed along the corridor, and the network-wide queue length. Shaded backgrounds indicate different demand levels—green, yellow, and red correspond to low, medium, and high demand periods, respectively. Note that these levels represent the rate of new vehicle generation, which may lag behind actual traffic conditions due to system inertia.

At the start of each high-level control cycle, the curves display abrupt transitions due to the 600-second measurement phase and policy activation. Group 1 and Group 2 prioritize corridor efficiency, resulting in smoother flow and fewer stops along the corridor but increased congestion across the network. Group 3 yields more balanced control by focusing on global performance. Among them, Group 2 achieves the most favorable trade-off, preserving corridor efficiency while limiting excessive queuing during peak hours. These results demonstrate the adaptability of the HLC, which effectively tailors coordination strategies to the configured objectives, enabling flexible management of complex urban traffic.

\section{Conclusion}
\label{sec:conclusion}
This study presents a hierarchical traffic signal control framework that integrates model-based coordination with reinforcement learning to meet the complex control demands of urban corridor management. By decomposing the control task into three interactive components—the High-Level Coordinator (HLC), Corridor Coordinator, and Hybrid Signal Agents (HSAs)—the proposed system achieves a balance between global coordination and local adaptability.

Specifically, we extend the max-flow coordination model for real-time deployment, introducing an MILP-based approach that optimizes green splits and offsets under high-demand conditions. We also design a learning-based signal agent architecture that supports constrained policy execution via action masking, ensuring that RL agents operate within the bounds of coordination strategies. The HLC further enhances system flexibility by dynamically assigning coordination modes based on predicted demand and corridor conditions.

Simulation results demonstrate the complementary strengths of the three strategies: MFC excels in maximizing throughput under oversaturated conditions, GWC minimizes corridor stops and maintains smooth progression, and PAC provides flexibility and responsiveness in light-demand scenarios. The hierarchical control design enables the system to exploit these advantages dynamically, adapting to varying traffic states and management goals.

Future work will explore extending this framework to larger networks with multiple corridors and incorporating uncertainty-aware decision-making to further improve robustness. In addition, real-world deployment studies and integration with connected vehicle data remain promising directions for validating and scaling the proposed approach.

\bibliography{mybib}

\end{document}